\documentstyle[epsf]{mn}
\newcommand{\m}{\rlap{$^{\dagger}$}}
\newcommand{\mm}{\rlap{$^{\ddagger}$}}
\newcommand{\coli}{{\small \jk--[12]}}
\newcommand{\colii}{{\small [12]--[25]}}
\newcommand{\coliii}{{\small [25]--[60]}}
\newcommand{\md}{\mbox{\rm d}}
\newcommand{\dm}{\relax\ifmmode{\dot{M}}\else{$\dot{M}$}\fi}
\newcommand{\teff}{\relax\ifmmode{T_{\rm eff}}\else{$T_{\rm eff}$}\fi}
\newcommand{\mete}{$M_{\rm env}$--$T_{\rm eff}$}
\newcommand{\pta}{\relax\ifmmode{P_{\rm a}}\else{$P_{\rm a}$}\fi}
\newcommand{\ptb}{\relax\ifmmode{P_{\rm b}}\else{$P_{\rm b}$}\fi}
\newcommand{\ha}{\relax\ifmmode{\rm H\alpha}\else{\rm H$\alpha$}\fi}
\newcommand{\mic}{\relax\ifmmode{\mu{\rm m}}\else{$\mu$m}\fi}
\newcommand{\jv}{{\rm\sl V}}
\newcommand{\jk}{{\rm\sl K}}

\newcommand{\zm}{\relax\ifmmode{\rm M_{\odot}}\else {M$_{\odot}$}\fi}
\newcommand{\zl}{\relax\ifmmode{{\rm L}_{\odot}}\else{L$_{\odot}$}\fi}
\newcommand{\zs}{\relax\ifmmode{{\rm R}_{\odot}}\else{R$_{\odot}$}\fi}
\newcommand{\kms}{\relax\ifmmode{\rm km\,s^{-1}}\else{km\,s$^{-1}$}\fi}
\newcommand{\ky}{\relax\ifmmode{\rm K\,yr^{-1}}\else{K\,yr$^{-1}$}\fi}
\newcommand{\zmy}{\relax\ifmmode{\rm\zm\,yr^{-1}}\else{\zm\,yr$^{-1}$}\fi}
\newcommand{\xt}[1]{\mbox{$\times 10^{#1}$}}
\newcommand{\eq}{equation}
\newcommand{\req}[1]{\eq~(\ref{#1})}
\newcommand{\sct}{Section}
\newcommand{\fig}{Fig.}
\newcommand{\figs}{Figs.}
\newcommand{\tbl}{Table}
\newcommand{\col}{colour}
\newcommand{\cols}{colours}
\newcommand{\cc}{colour-colour}
\newcommand{\beh}{behaviour}
\newcommand{\cen}{centre}
\newcommand{\lbd}{labelled}
\newcommand{\q}{}
\newcommand{\qs}{,}
\newcommand{\rc}{,}
\newcommand{\iras}{{\it IRAS}}
\newcommand{\iso}{{\it ISO}}
\newcommand{\cld}{{\sc cloudy}}
\begin{document}
\title{The spectral evolution of post-AGB stars}
\author[P.A.M. van Hoof, R.D. Oudmaijer and L.B.F.M. Waters]
{P.A.M. van Hoof,$^{1}$
R.D. Oudmaijer$^{2,1}$ and
L.B.F.M. Waters$^{3,4}$\\
$^1$Kapteyn Astronomical Institute, P.O. Box 800, 9700 AV Groningen,
The Netherlands\\
$^2$Imperial College, Blackett Laboratory, c/o Astronomy Group, Dept. of
Physics, London SW7 2BZ, United Kingdom\\
$^3$Astronomical Institute ``Anton Pannekoek'', University of Amsterdam,
Kruislaan 403, 1098 SJ Amsterdam, The Netherlands\\
$^4$SRON Laboratory for Space Research, P.O. Box 800, 9700 AV Groningen,
The Netherlands
}
\date{received, accepted}
\maketitle

\begin{abstract}
A parameter study of the spectral evolution of a typical post-AGB star,
with particular emphasis on the evolution of the infrared \cols, is
presented.  The models
are based on the latest evolutionary tracks for hydrogen burning post-AGB
stars.
For such tracks the
evolutionary rate is very dependent on the assumed mass loss rate as a
function of time.
We investigate this effect by modifying the mass loss prescription.  The
newly calculated evolutionary rates and density distributions are used to
model the spectral evolution of a post-AGB star with the photo-ionization
code \cld, including dust in the radiative transfer.  Different assumptions
for the dust properties and dust formation are considered.  It is shown
that by varying these parameters in a reasonable way, entirely different
paths are followed in the \iras\ \cc\ diagram.  First of all, the effects
of the evolution of the central star on the expanding dust shell can not be
neglected.
Also the dust properties and the definition of the end of the AGB phase
have an important effect.
The model tracks show that objects occupying the same location in the
\iras\ \cc\ diagram can have a different evolutionary past, and therefore
the position in the \iras\ \cc\ diagram alone can not {\em a priori} give a
unique determination of the evolutionary status of an object.
An alternative \cc\ diagram, the \coli\ vs.\ \colii\ diagram, is
presented. The tracks in this diagram seem less affected by particulars of
the grain emission. This diagram may be a valuable additional tool for
studying post-AGB evolution.
\end{abstract}


\section{Introduction}

The transition phase between the Asymptotic Giant Branch (AGB) and
planetary nebulae (PNe) has gained much attention over the last decade.
AGB stars lose mass fast and get obscured by their circumstellar
dust.  As the star leaves the AGB, its mass loss rate decreases
significantly and the star may become sufficiently hot to ionize its
circumstellar material and be observable as a PN.  During the
transition from the AGB to the PN phase (the post-AGB or proto-planetary
nebula phase) the dust shell created during the AGB moves away from the
central star and becomes optically thin after a few hundred years; the
obscured star becomes observable.  The transition time from the AGB to the
PN phase is estimated to be a few thousand years (e.g.\ Pottasch\q\ 1984).

Whereas PN are relatively easy to find because of their rich optical
emission line spectra, post-AGB stars have more inconspicuous spectra
and are therefore much harder to find.  The number of post-AGB stars only
started to become large after the \iras\ mission, that was successful in
detecting objects surrounded by circumstellar dust.

Several samples of post-AGB stars are presented in the literature (e.g.\
Volk \& Kwok\q\ 1989\qs\ Hrivnak, Kwok \& Volk\q\ 1989\qs\ van der Veen,
Habing \& Geballe\q\ 1989\qs\ Trams et al.\q\ 1991\qs\ Oudmaijer et al.\q\
1992\qs\ Slijkhuis\q\ 1992).  Most of these objects are stars with
supergiant-type spectra, surrounded by dust shells.  These samples of
post-AGB stars have in common that the original criteria which were
employed to find them implicitly made assumptions on the spectral energy
distribution (SED) of post-AGB stars.  Some authors used criteria on the
\iras\ \cols, because post-AGB stars were expected to be located in a
region in the \iras\ \cc\ diagram between AGB stars and PN (e.g.\ Volk \&
Kwok\q\ 1989\qs\ Hrivnak et al.\q\ 1989\qs\ van der Veen et al.\q\ 1989\qs\
Slijkhuis\q\ 1992).  Other authors loosened this criterion and searched for
objects in the entire \cc\ diagram, but with an additional criterion that
the central star should be optically visible (e.g.\ Trams et al.\q\
1991\qs\ Oudmaijer et al.\q\ 1992\qs\ Oudmaijer\q\ 1996).

Such samples are subject to selection effects, so the objects that have
been selected do not have to be representative for the entire population of
post-AGB stars.  In order to understand these selection effects and to
obtain a handle on the kinds of objects that could have been missed, it is
useful to investigate the spectral energy distribution from a theoretical
point of view by following the spectral evolution of a post-AGB star with
an expanding circumstellar shell.  Moreover, this type of study allows one
to investigate and understand the processes that occur in the circumstellar
shell during the transition.

Several such studies have been published.  Most of these studies focus on
the expanding dust shell with a dust radiative transfer model.  Authors
like e.g.  Siebenmorgen, Zijl\-stra \& Kr\"ugel (1994), Szczerba \& Marten
(1993), Loup (1991), Slijkhuis \& Groenewegen (1992) and Volk \& Kwok
(1989) performed calculations describing the evolution of the circumstellar
dust shell.  Work concerning the evolution of a star with an expanding
shell has also been performed with photo-ionization codes (Volk\q\ 1992),
and hydrodynamical models (Frank et al.\q\ 1993\qs\ Mellema\q\ 1993\qs\
Marten \& Sch\"onberner\q\ 1991).  None of these models include dust,
except for Volk (1992) who used the output of the photo-ionization code
\cld\ (Ferland\q\ 1993) as input for a dust model.

Our objective is to investigate the spectral evolution of a hydrogen burning
post-AGB star
with a photo-ionization model containing a dust code.  The aim of this
work is twofold: \\
Firstly we investigate the processes in the circumstellar envelope.  The
emphasis in this paper will be on the infrared properties of this shell.
For this, both the expansion of the shell and the evolution of the central
star have to be taken into account. \\
Secondly, we investigate the influence of certain assumptions on the
evolutionary timescales.  Only few post-AGB evolutionary grids have been
published, never giving a fine grid for the coolest part of the evolution.
The original Sch\"onberner tracks (1979, 1983) presented only a limited
number of time points of the evolution, while Vassiliadis \& Wood (1994)
omit the phase between 5000~K and 10\,000~K altogether.  The predicted
timescales that are available are calculated with a pre-defined end of the
AGB and assumed post-AGB mass loss rates.
These choices can influence the post-AGB evolutionary timescales
considerably, as was already demonstrated by Trams et al. (1989) and
G\'orny, Tylenda \& Szczerba (1994).
The situation has changed now with the results of Bl\"ocker
(1995a,b), who calculated new evolutionary sequences, and made extensive
tables available describing certain key parameters during the post-AGB
phase.  The published relation between the envelope mass and effective
temperature allows one to construct detailed timescales using one's own mass
loss prescriptions during the (post-) AGB evolution, which makes the model
results less dependent on the mass loss formulation that was used by
Sch\"onberner (1979, 1981, 1983).

In this paper first we describe the method to calculate evolutionary
timescales and the adopted mass loss prescriptions.  The results of
Bl\"ocker (1995a,b) are used as a basis to create synthetic evolutionary
tracks, and some aspects of the evolutionary timescales that are predicted
are discussed.  Next we describe the photo-ionization code \cld\ that was
used and the assumptions that were made to conduct the study of the
spectral evolution of post-AGB stars. We then present the first results of
a parameter study of a typical post-AGB object, based on the 0.605~\zm\
track from Bl\"ocker (1995b).

\section{The central star evolution}

Many stellar evolutionary models are presented in the recent literature,
but only two groups calculate the AGB quantitatively including mass loss.
These are Vassiliadis \& Wood (1993, 1994) and Bl\"ocker \& Sch\"onberner
(1991) and Bl\"ocker (1995a,b).  Both groups calculate the evolution of a
star from the main sequence through the red giant phase to the white dwarf
stage.  There are slight differences in the core mass -- luminosity
relations and the use of a different initial -- final mass relation, but
their results show qualitatively the same \beh\ in the evolution of stars
in the HR diagram.  The main differences between the models are the mass
loss prescriptions on the AGB.

The AGB mass loss rates in the formulation of Vassiliadis \& Wood are
derived from the mass loss -- pulsation period (\dm--$P$) relation given by
Wood (1990).  For periods longer than 500~d for low mass stars, and larger
pulsation periods for higher mass objects, a maximum value of the AGB mass
loss rate is invoked (of the order of 10$^{-5}$~\zmy).  Bl\"ocker used a mass
loss rate which is dependent on the luminosity of the star.  He fitted the
results of Bowen's (1988) theoretical study of mass loss in Mira variables.
Basically this is the Reimers mass loss (Reimers\q\ 1975) multiplied by a
luminosity dependent factor. These mass loss rates are not limited by a
maximum value.  Since the adopted AGB mass loss rates by the above authors
differ, some striking differences between the results of their calculations
exist.  The \dm--$P$ relation of Wood (1990) is questioned by Groenewegen
\& de Jong (1994).  Using the synthetic evolutionary model of Groenewegen
\& de Jong (1993), they were able to fit the luminosity function of carbon
stars in the LMC with the Bowen mass loss adopted by Bl\"ocker (1995a), but
not with the Vassiliadis \& Wood mass loss rates.

The mass loss rates do not only govern the evolution on the AGB.  During
the post-AGB phase, the mass loss rates also have a drastic influence on
the timescales of the AGB~-- PN transition.  Since the temperature of the
star for a given core mass is determined by the mass of the stellar
envelope, larger mass loss rates will cause the star to evolve to higher
temperatures more quickly and can therefore decrease the timescale of the
transition strongly.  Trams et al.\ (1989) showed that when the adopted
post-AGB mass loss rates of a 0.546~\zm\ star are raised by a factor of 5
to 10 to a value of 10$^{-7}$~\zmy, the transition time from the AGB to the
PN phase is shortened from 100\,000~yr to only 5000~yr.  This would make a
low mass star readily observable as a PN, while the (longer) predicted
timescales prevent such objects to become an observable PN since the
circumstellar shell would have moved far away from the star and would have
dispersed into the interstellar medium long before the star emits a
sufficient amount of ionizing photons.  In order to study the evolutionary
timescales of post-AGB stars, it is important to understand post-AGB mass
loss better.

\subsection{Determination of the evolutionary timescales}

The evolutionary timescales of hydrogen burning post-AGB stars
are determined by the core mass and the mass
loss rates of the star.  In this section we present the computational
details of this procedure, compute the evolutionary timescales and
investigate several aspects of these timescales.  We use the tracks by
Bl\"ocker (1995a,b) to determine the evolutionary timescales.  Bl\"ocker
calculated several tracks for his investigation of the evolution of stars
on the AGB and beyond.  The tracks include a complete calculation of a
3~\zm\ object that ends as a 0.605~\zm\ white dwarf and a 4~\zm\
(0.696~\zm)
sequence.  To these  sequences, the
1.0~\zm\ (0.565~\zm) track of Sch\"onberner (1983) was added.  We start
with a recapitulation of the mass loss laws we have used (see also
Bl\"ocker\q\ 1995b).

\subsubsection{The mass loss laws}

During the red giant branch, the mass loss is parameterized by the Reimers
mass loss rate $\dm_{\rm R}$ given in \req{reimers}, where the scaling
parameter $\eta$ is set to 1 for all tracks.

\begin{equation}
\dm_{\rm R}/(\zmy) = 4 \times 10^{-13}\,\eta\,\frac{(L/\zl)(R/\zs)}{(M/\zm)}
\label{reimers}
\end{equation}
With $L$, $R$ and $M$ the luminosity, radius and mass of the star.  Since
AGB stars lose mass at a faster rate than red giants, a different
formulation is adopted for the AGB mass loss rates.  To this end, Bl\"ocker
fitted the numerical results of Bowen, as given in \req{bowen}.

\begin{equation}
\dm_{\rm B1}/(\zmy) =
\label{bowen}
\end{equation}
$\hfill 4.83\times10^{-9}\,(M_{\rm ZAMS}/\zm)^{-2.1}(L/\zl)^{2.7}\dm_{\rm R}$
\[
\dm_{\rm B2}/(\zmy) =
\hfill 4.83\times10^{-9}\,(M/\zm)^{-2.1}(L/\zl)^{2.7}\dm_{\rm R} \]
The difference between $\dm_{\rm B1}$ and $\dm_{\rm
B2}$ is the division by the current mass of the star instead of the initial
mass of the star.  The timescales for the 0.565~\zm\ track of Sch\"onberner
were recalculated by us using the \mete\ relation given by Sch\"onberner
but with the mass loss prescriptions of Bl\"ocker.  For all tracks the
$\dm_{\rm B1}$ prescription was used, except for the 0.605~\zm\ track where
$\dm_{\rm B2}$ was used, which results in larger AGB mass loss rates.

The mass loss rates can reach values of the order of 10$^{-4}$~\zmy\ to
10$^{-3}$~\zmy\ and it is clear that if this mass loss would last for a
long time, the star would evaporate.  Thus, an end to the AGB (-wind) has to
be invoked.
Bl\"ocker used the pulsation period to define the end of the high mass loss
phase; he assumed that AGB stars pulsate in the fundamental mode, where the
period can be calculated using \req{pulper} (Ostlie \& Cox\q\ 1986).
\begin{equation}
\lg(P_0/{\rm d})  = -1.92 - 0.73\lg(M/\zm) + 1.86\lg(R/\zs)
\label{pulper}  
\end{equation}
When the central star has reached an inferred pulsation period of \pta\ = 100~d
(which occurs when the star has a surface temperature somewhere between
roughly 4500~K and 6000~K) the Bowen mass loss stops.  When the star has
subsequently reached an inferred pulsation period of \ptb\ = 50~d the post-AGB
mass loss starts; in between the mass loss rates are connected by a smooth
transition. Hence the following definitions will be used in the remainder of
the paper: the AGB phase is that part of the evolution where the pulsation
period is greater than \ptb, the (AGB) transition phase is that part of the
evolution where the pulsation period is in between \pta\ and \ptb, and finally
the post-AGB phase is that part of the evolution where the pulsation period is
smaller than \ptb.

One should realize that post-AGB stars with pulsation periods larger than
50~d exist (e.g. HD 52961 with a period of 72~d; Waelkens et al.\q\ 1991\qs\
Fernie\q\ 1995).
Thus the pulsation period recipe used by Bl\"ocker should only be regarded
as an approximate parameterization of the end of the AGB.

For lower temperatures the post-AGB mass loss is given by the Reimers law
(\ref{reimers}).  Since the Reimers mass loss is proportional to
\teff$^{-2}$ for a constant luminosity and stellar mass, the post-AGB mass
loss rates decrease during the evolution of the object.  A radiation driven
wind will take over when the star has reached temperatures above
approximately 20\,000~K.  The mass loss rate for this wind, based on
Pauldrach et al.\ (1988), is given by \req{mcpn}.  Hence at any stage of
the post-AGB evolution either \req{reimers} or (\ref{mcpn}) is used,
whichever of the two yields the biggest mass loss.

\begin{equation}
\dm_{\rm CPN}/(\zmy) = 1.29 \times 10^{-15}\,(L/\zl)^{1.86}
\label{mcpn}
\end{equation}

During the evolution on the AGB and beyond, the envelope mass is reduced
due to two processes.  In the outer parts of the envelope, mass is lost
through a wind, and at the bottom of the envelope, mass is diminished by
hydrogen burning.  From Trams et al.\ (1989) we find the mass loss due to
hydrogen burning:

\begin{equation}
\dm_{\rm H}/(\zmy) = 1.012 \times 10^{-11}\,(L/\zl)X_{\rm e}^{-1}
\label{hburn}
\end{equation}
With $X_{\rm e}$ the hydrogen mass fraction in the envelope (70~\%).

\subsubsection{The evolutionary timescales}
\label{trans:tim}

The evolutionary timescales for hydrogen burning post-AGB stars depend
firstly on the core mass and secondly on
the mass loss rates.  The evolutionary timescales can be calculated
relatively easy by making use of the fact that there exists a unique
relation between the envelope mass and the stellar temperature for every
core mass.  One can calculate the evolutionary timescales by combining this
relation with the mass loss prescriptions given in the previous section.
Dr.\ Bl\"ocker kindly provided us with tables from which the \mete\
relation could be reproduced.

The timescale for the envelope depletion is given by:
\begin{equation}
\frac{\Delta M_{\rm env}}{\Delta t}  = \dm_{\rm wind} + \dm_{\rm H}
\end{equation}
Using the \mete\ relation this expression can be transformed into an
expression for the evolutionary rate $\Delta T_{\rm eff}/\Delta t$ of the
central star
\begin{equation}
\frac{\Delta T_{\rm eff}}{\Delta t} =
\frac{\Delta T_{\rm eff}}{\Delta M_{\rm env}} \times
\frac{\Delta M_{\rm env}}{\Delta t} =
\frac{\dm_{\rm wind} + \dm_{\rm H}}{\frac{\md M_{\rm env}}{\md T_{\rm eff}}}
\label{rate}
\end{equation}
Integrating this expression yields the actual timescales for the evolution
of the central star.

\begin{figure}
\begin{center}
\setlength{\unitlength}{1in}
\mbox{\epsfxsize=0.45\textwidth\epsfbox[20 145 456 686]{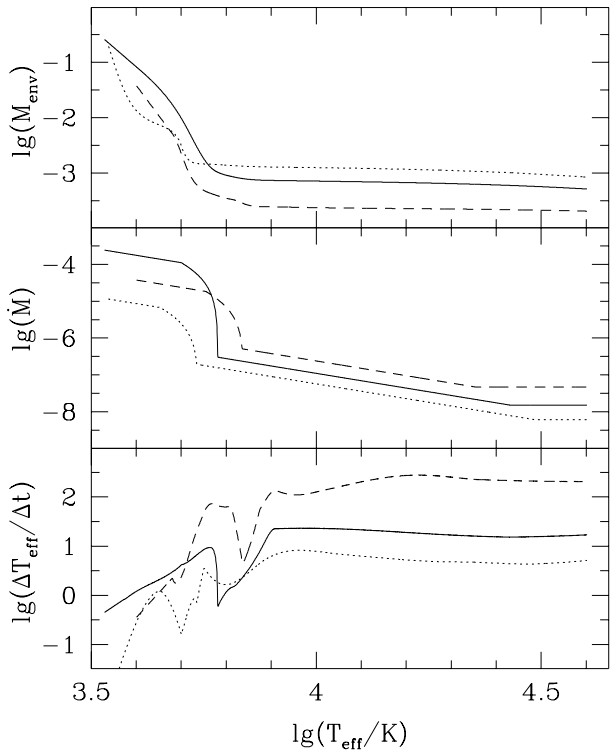}}
\caption[The \mete\ relations, mass loss rates and evolutionary rates]
{The \mete\ relations, mass loss rates and resulting evolutionary rates for
the 0.565~\zm\ (dotted line), 0.605~\zm\ (solid line) and 0.696~\zm\
(dashed line) tracks.  The envelope mass is given in solar masses, the mass
loss rate in solar masses per year and the evolutionary rate in kelvin per
year.  Note that the 0.605~\zm\ track has a larger AGB mass loss rate. For
this particular track, the B2 mass loss rate was used.}
\label{plot5}
\end{center}
\end{figure}

In \fig~\ref{plot5} some relations are presented for the 0.565~\zm,
0.605~\zm\ and 0.696~\zm\ tracks.  The upper panel shows the envelope mass
as function of photospheric temperature.
The second panel presents the mass loss rates as function of temperature,
and the third panel shows the evolutionary rate in kelvin per year as
function of temperature.  In the second panel it is visible that the
transition phase occurs at higher temperatures for larger core masses with
the recipes described above.

The evolutionary rate, mass loss rate and envelope mass are related to each
other in the following way.  The evolutionary rate in terms of increase in
temperature per year is slow on the steep part of the \mete\ relation, and
more rapid on the shallow part.  Larger mass loss rates imply of course more
rapid changes in the stellar temperature.  These effects are visible in
\fig~\ref{plot5}: at low temperatures in the post-AGB phase the evolutionary
rates
are smallest, while for higher temperatures the evolutionary rate
increases.  The minima around lg(\teff/K)~$\approx$ 3.7 to 3.9 correspond
to the onset of the smaller post-AGB mass loss, slowing down the evolution
which accelerates later on the shallow part of the \mete\ relation.

Interestingly, the minimum in the evolutionary rate occurs right after the
end of the AGB transition phase.  The net increase in effective temperature
in kelvin per year is the slowest for all tracks just after the start of
the post-AGB evolution, which is when the temperature of the objects
correspond to G or F spectral types.  The increase in temperature is less
than 1~\ky\ for the 0.565~\zm\ and 0.605~\zm\ tracks.  How does this
evolutionary rate compare with the observations? Fernie \& Sasselov (1989)
calculated the possible increase in temperature for UU Her stars.  From the
absence of a change in pulsation period of UU Her, 89 Her and HD 161796,
they place an upper limit of 0.5~\ky\ on the temperature increase.  Their
conclusion was that these objects can not be post-AGB stars because the
evolutionary rates should be much higher.  However, this may be the case
when one averages over the entire post-AGB temperature span, but the
observed lack of temperature increase is
consistent with the predictions for this temperature range,
as was already shown by Sch\"onberner \& Bl\"ocker (1993).  Therefore a
post-AGB nature for the UU Her stars can not be excluded on this basis.

\subsection{The influence of mass loss on the evolutionary timescales}

It is evident from the above that the value of the post-AGB mass loss rate
has an important effect on the evolutionary timescales.  However, the
mass loss rate is not known observationally for cool post-AGB stars.  The
usual tracer of mass loss, \ha\ emission, which is often observed in the
spectra of post-AGB objects is likely to be the result of stellar
pulsations (see the discussion by Oudmaijer \& Bakker\q\ 1994 and L\`ebre
et al.\q\ 1996).  A possible
tracer of mass loss in cool post-AGB stars is the CO first-overtone
emission at 2.3~\mic\ (Oudmaijer et al.\q\ 1995), but, given this is true,
the mass loss rates still have to be determined.

The lack of theoretical and observational values for post-AGB mass loss
rates forced Sch\"onberner and Bl\"ocker to resort to the heuristic Reimers
law for the cool part of the post-AGB evolution.  As an illustration of the
effect of the post-AGB mass loss rates on the evolutionary timescales, we
have calculated these timescales for three core masses with the
post-AGB mass loss rate at 0, 1, 5 and 10 times the standard post-AGB value
(indicated as 0$\times$pAGB etc.). The
results for the 0.565~\zm, 0.605~\zm\ and 0.696~\zm\ tracks are presented
in \tbl~\ref{time565}, where the timescales since the end of the transition
mass loss phase are given.  The increase of the post-AGB mass loss rates
indeed decreases the timescales of the evolution, confirming the results of
Trams et al.\ (1989) and G\'orny et al. (1994).
The 0$\times$pAGB mass loss rate effectively
determines the slowest possible evolutionary speed, since the only mass
loss is through hydrogen burning. On average, the difference in speed
between 0$\times$pAGB and 1$\times$pAGB is roughly a factor of two to
three.

\begin{table}
\caption[Evolutionary timescales for different values of post-AGB mass loss
rates]
{Evolutionary timescales for different values of post-AGB mass loss
rates; the post-AGB phase starts at \ptb\ = 50~d.}
\label{time565}
\begin{footnotesize}
\begin{tabular}{rrrrrrrrrrrrrr}
\hline
$\dm_{\rm PAGB} $   & {\scriptsize 0$\times$pAGB}   &  {\scriptsize 1$\times$pAGB}   & {\scriptsize 5$\times$pAGB} &
{\scriptsize 10$\times$pAGB} \\
 \teff\ (K) \\
\hline
\multicolumn{5}{c}{$M_{\rm core}$ = 0.565~\zm, $L$ = 3891~\zl} \\
\hline
      5\,426 &         0 &       0 &       0 &       0 \\
      5\,500 &       312 &      76 &      18 &       9 \\
      6\,000 &      1113 &     286 &      70 &      36 \\
      7\,000 &      2810 &     815 &     207 &     107 \\
     10\,000 &      3974 &    1306 &     349 &     182 \\
     15\,000 &      5172 &    2092 &     651 &     353 \\
     20\,000 &      6383 &    3089 &    1141 &     653 \\
     25\,000 &      7544 &    4156 &    1768 &    1068 \\
     30\,000 &      8709 &    5292 &    2526 &    1604 \\
\hline
\multicolumn{5}{c}{$M_{\rm core}$ = 0.605~\zm, $L$ = 6310~\zl} \\
\hline
      6\,042 &         0 &       0 &       0 &       0 \\
      6\,500 &      1671 &     458 &     113 &      58 \\
      7\,000 &      2510 &     715 &     178 &      92 \\
     10\,000 &      3122 &     948 &     241 &     125 \\
     15\,000 &      3487 &    1181 &     325 &     171 \\
     20\,000 &      3822 &    1458 &     452 &     248 \\
     25\,000 &      4154 &    1769 &     625 &     358 \\
     30\,000 &      4481 &    2093 &     832 &     501 \\
\hline
\multicolumn{5}{c}{$M_{\rm core}$ = 0.696~\zm, $L$ = 11\,610~\zl} \\
\hline
      6\,846 &         0 &       0 &       0 &       0 \\
      7\,000 &        94 &      29 &       7 &       4 \\
      7\,500 &       192 &      62 &      15 &       8 \\
     10\,000 &       245 &      84 &      22 &      11 \\
     15\,000 &       287 &     111 &      30 &      16 \\
     20\,000 &       309 &     129 &      38 &      21 \\
     25\,000 &       331 &     151 &      50 &      28 \\
     30\,000 &       354 &     174 &      65 &      38 \\
\hline
\end{tabular}
\, \\
\end{footnotesize}
\end{table}

\subsection{Distribution over spectral type} 
\label{histosec} 

The availability of the evolutionary rates allows us to investigate the
predicted distribution over spectral type.  Oud\-maijer, Waters \& Pottasch
(1993) and Oudmaijer (1996) used the coarse grid of the Sch\"onberner
(1979, 1983) tracks and found that a star spends by far most of the
time as a B-type star.  Only for the 0.644~\zm\ track half of
the time is spent in the G phase, and somewhat less as a B star, while
almost no time is spent as an F or A star.

One might not expect that an object would spend a large fraction of the
time as a B star, because the evolutionary rates are largest for B spectral
type (\fig~\ref{plot5}).  This can be understood however when we consider
the large range of temperatures that corresponds to spectral type B: roughly
between 10\,000~K and 30\,000~K.  In contrast, A stars only have a
temperature range between approximately 7500~K and 10\,000~K.  The large
evolutionary rates multiplied by the temperature range then result in a
longer time spent as a B star during the post-AGB evolution.  The large
fraction of time that is spent as a G star in the 0.644~\zm\ track is
explained by the steep \mete\ relation for temperatures less than
approximately 6000~K, as can be deduced from \req{rate}.  The distribution
over spectral type for the tracks presented here with 1$\times$pAGB
mass loss is calculated using the conversion from effective temperature to
spectral type listed by Strai\v{z}ys \& Kuriliene (1981).  These are
\teff(A0I) = 9800~K, \teff(F0I) = 7400~K and \teff(G0I) = 5700~K.

\begin{figure}
\begin{center}
\mbox{\epsfxsize=0.45\textwidth\epsfbox[20 202  552 641]{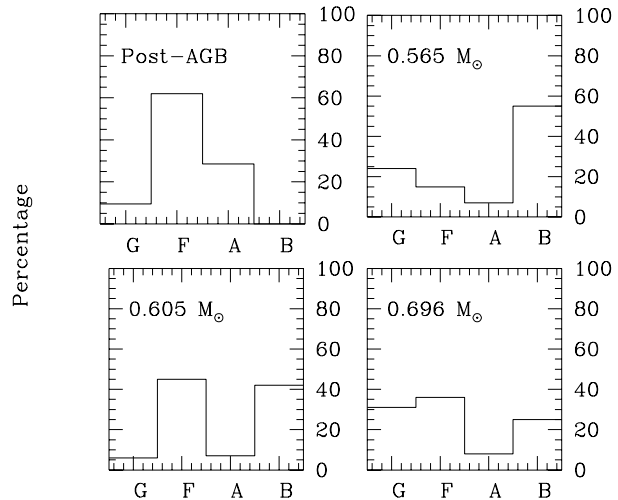}}
\caption[Comparison of the distribution over spectral type for certain
model tracks with observations]
{Distribution over spectral type. Upper left panel; the post-AGB candidate
stars in the sample of Oudmaijer et al.\ (1992), the other panels show the
predicted distribution for the 0.565~\zm, 0.605~\zm\ and 0.696~\zm\ tracks.
}
\label{histo}
\end{center}
\end{figure}

To allow for a comparison with the distributions presented by Oudmaijer et
al.\ (1993) we will assume in this section that the post-AGB phase starts when
\teff\ = 5000~K and ends when \teff\ = 25\,000~K.  The distributions can be
easily obtained by calculating the time spent as a G star (5000~K -- G0), F
star (G0 -- F0) etc., and subsequently dividing these numbers by the total
time spent as a post-AGB star.
The resulting distributions
are plotted in \fig~\ref{histo}.
For comparison the distribution over spectral type of the sample of
21 post-AGB objects in the list of Oudmaijer et al.\ (1992) is given.  The
observed distribution peaks at F, while no B stars are found at all.%
\footnote{
The few B type objects in the sample of Oudmaijer et al.\
(1992) appear to have low effective temperatures. HR 4049 and HD 44179 (the
central star of the Red Rectangle) are listed in the literature as a B
star, but abundance analyses showed that the effective temperature
is lower than typical for a B star: about 7500~K (late A or early F type;
van Hoof et al.\q\ 1991\qs\ Waelkens et al.\q\ 1992).  The reason that
these objects were classified as B, instead of late A stars,
is the extreme metal deficiency of these stars, so that
few metallic lines are present in the spectra. Thus the observed spectra
mimic the spectra of hot objects.  }

One should realize that planetary nebula central stars with temperatures
below 25\,000~K are observed: e.g. IRAS19336--0400 with \teff~= 23\,000~K
(Van de Steene, Jacoby \& Pottasch\q\ 1996\qs\ Van de Steene \& van Hoof\q\
1995).
Hence one could also assume an upper limit of
20\,000~K for the post-AGB regime.
However, the distribution over spectral type is not very different
in this case, the fraction of B-type stars
will be lower by approximately 10~\%.

The 0.605~\zm\ distribution is different from the results presented by
Oudmaijer et al.\ (1993) for the 0.598~\zm\ Sch\"onberner track.
In the present plot, the distribution peaks at F, while the 0.598~\zm\
distribution peaks at G.  The distribution for the same track is different
in Oudmaijer (1996), since there the evolutionary timescales were shortened
by 1000~yr `in order to make the old 0.598~\zm\ track consistent with the
new calculations' (Marten \& Sch\"onberner\q\ 1991).  This resulted in a
distribution that strongly peaks at spectral type~B.

The main differences between the old 0.598~\zm\  and new 0.605~\zm\ 
calculations are the progenitor mass of the star (1~\zm\ and 3~\zm\
respectively), the mass loss prescriptions and the definition of the end of
the AGB.  In the Sch\"onberner calculations, no transition wind was assumed:
the AGB mass loss would abruptly change into a post-AGB wind at 5000~K.
In the Bl\"ocker calculations, a transition wind is assumed between 5000~K and
6000~K for the 0.605~\zm\ track.  This implies {\em shorter} timescales for the
G type phase of the 0.605~\zm\ track with respect to the older calculations,
but does not affect the time spent as F, A or B stars,
where for both tracks a Reimers post-AGB mass loss is assumed.  The only
explanation we can find for the apparent difference in evolutionary speed
is that the dependence of the temperature on the
envelope mass of the 0.605~\zm\ track is different from the 0.598~\zm\ track.
Apparently, the slope of the \mete\ relation can differ significantly for
stars with a different evolutionary past, even when the resulting core masses
are nearly identical.

The predicted distributions in \fig~\ref{histo} show large differences.
For the smallest core mass, most
of the time is spent as a B star.  For the 0.605~\zm\
case the lifetime is distributed evenly over the F and B
phase. The fastest track with a core mass of 0.696~\zm\ shows a minimum at
A, while the rest of the lifetime is spread evenly over G, F and~B.
The 0.605~\zm\ distribution nicely reproduces the observed peak at F.
However, all tracks predict many more B stars than observed.

In order to compare the observed and predicted distributions more
quantitatively, we have performed a Kolmogorov-Smirnov test using both an upper
limit of 20\,000~K and 25\,000~K for the post-AGB regime. The resulting
probabilities are given in \tbl~\ref{ks:tab}. From these
tests it appears that 0.565~\zm\ track is an unlikely model for the observed
post-AGB sample. On the other hand, the 0.605~\zm\ and the 0.696~\zm\ tracks
can not be excluded. Given the fact
that stars on the 0.696~\zm\ track evolve much faster than on the 0.605~\zm\
track, they would constitute only a small fraction of the total number of
observable post-AGB stars.
We will adopt the 0.605~\zm\ track to describe the post-AGB
evolution in the remainder of this paper.

\begin{table}
\caption[The results of the Kolmogorov-Smirnov test]
{The results of the Kolmogorov-Smirnov test comparing various evolutionary
tracks with the observed post-AGB sample. The tests have been performed using
a range of temperatures for the post-AGB regime of 5000~K to 20\,000~K
(second column) or alternatively 5000~K to 25\,000~K (third column).}
\label{ks:tab}
\begin{tabular}{rrr}
\hline
track &   probability &   probability \\
  \zm &     20\,000~K &     25\,000~K \\
\hline
0.565 &    4.5\xt{-4} &    2.8\xt{-6} \\
0.605 &    2.4\xt{-2} &    6.4\xt{-4} \\
0.696 &    6.7\xt{-2} &    1.3\xt{-1} \\
\hline
\end{tabular}
\end{table}

This exercise shows that the predictions of the distribution over spectral
type are subject to large uncertainties, depending both on the
\mete\ relation and the assumed mass loss prescription.
However, regardless what assumptions are made,
one would expect a fair number of B-type post-AGB stars,
which are not
observed in the sample depicted in \fig~\ref{histo}.
Hence this discrepancy remains unresolved.
Reversely it can be stated that when a larger sample of post-AGB stars with
reliable temperature determinations becomes available, the procedure described
here can be a very effective means for testing evolutionary tracks.

\section{The model}
\label{model}

In order to calculate the spectral evolution of post-AGB stars, we used the
photo-ionization code \cld\ version 84.12a (Ferland\q\ 1993).  Some
modifications have been made to the code to facilitate the computations.
The most important change was
the introduction of several broadband photometric filters, including the
Johnson and the \iras\ filters. The in-band fluxes for these filters were
calculated by folding both the spectral energy distribution and the
emission line contribution with the filter passband.  Internal extinction
due to continuum opacities was included both in the continuum and the line
contribution.

A dust model written by P.G. Martin was already included in the original
code.  For our modeling we used the grain species \lbd\ `ISM Silicate'
and `ISM Graphite'.  The optical constants were taken from Martin \&
Rouleau (1991).  The absorption and scattering cross sections were
calculated assuming a standard ISM grain size distribution (Mathis, Rumpl
\& Nordsieck\q\ 1977).  All calculations were done assuming a dust-to-gas
mass ratio of 1/150.  For the chemical composition of the gas we assumed
the abundances given in Aller \& Czyzak (1983), supplemented with educated
guesses for elements not listed therein (as given in \cld).

The original code only allowed for the computation of a model with a
constant dust-to-gas ratio throughout the entire nebula.  However, the
density profiles we calculate extend from the stellar surface outward.  It
is not realistic to assume that dust is present near the stellar surface
and therefore we introduced new code in \cld\ to solve this problem. This
enabled the dust to exist only outside a prescribed radius or,
alternatively, only in those regions where the equilibrium temperature of
the dust would be below a prescribed sublimation temperature. These
prescriptions work as a binary switch: at a certain radius either no dust
or the full amount is present. In those regions where dust exists, the
dust-to-gas ratio is assumed to be constant.

Two models for the dust formation are adopted. In the first model it is
assumed that dust is only formed in the AGB wind, hence
in material that was ejected before the stellar pulsation period reached \ptb.
We will call this the AGB-only dust formation model.  In the
second model it is assumed that the dust formation continues in the
post-AGB wind.
We will call this the post-AGB dust formation model.  Due to limitations of
the code, which we discuss below, we will only investigate the spectral
evolution after the post-AGB phase has started.  This implies that for the
AGB-only dust formation model, the inner dust radius is already at a
distance from the central star and the equilibrium temperature of the
grains is always below the sublimation temperature.  In the post-AGB dust
formation model we assume that dust only exists in those parts of the
nebula where the equilibrium temperature of the grains is below the
sublimation temperature.  The assumed values for the sublimation
temperature
are 1500~K for graphite and 1000~K for silicates.

It should be noted that we only try to model dust {\em formation} and not
the {\em destruction} of grains by the stellar UV field or shocks.
Especially in the AGB-only dust formation model the grains at the inner
dust radius are always exposed and it is expected that they eventually will
be destroyed.
However, little is known about grain destruction, and the rates at which
this destruction occurs are very uncertain.
In the case of continuing dust formation in the post-AGB wind this problem
can be expected to be of lesser importance since there is a constant supply
of new grains shielding the older grains.

The radiative transport in \cld\ is treated in one dimension only, i.e.\
the equations are solved radially outwards.
This assumption makes the code unsuitable to compute models of nebulae with
significant amounts of scattered light and/or diffuse emission when the nebula
has a moderate to high absorption optical depth.
For low optical depths re-absorption of diffuse emission in
the nebula is negligible and the assumptions in \cld\ work very well.
However, for moderate to high optical depths the re-absorption of diffuse
emission that is produced in the outer parts of the nebula and is radiated
inwards becomes important. The assumptions made in \cld\ make it impossible
to account for this energy source, nor for the amount of flux absorbed in
these regions.
Since the circumstellar envelope is optically thick in the AGB phase, our
calculations always start shortly after the transition from the AGB to the
post-AGB phase is complete.

For low temperature models the only source of diffuse light at optical and UV
wavelengths is scattering of
central star light by dust grains; for the highest temperature models
bound-free emission also plays a role. Since for the highest temperatures
the models have a low absorption optical depth,
the bound-free emission causes only minor problems
and we can judge the quality of the models primarily by
investigating the (wavelength-averaged) scattering optical depth. An example of
this approach will be shown in \sct~\ref{mod:runs}.

In the calculations, the central star was assumed to emit as a blackbody.
All the models were calculated for a distance of 1~kpc.

\subsection{Density profiles of the circumstellar shell}

The density profiles of the expanding shell are calculated for the
homogeneous and spherical case.  Using the mass loss prescriptions
described above and assuming an outflow velocity they can be calculated for
any moment in time $t$.
\begin{equation}
\rho(r,t)  = \frac{\dm_{\rm wind}(t_{\rm ej})}{4\pi r^{2}(t,t_{\rm ej})
v_{\rm exp}(t_{\rm ej})}
\end{equation}
with
\[ r(t,t_{\rm ej}) = R_{\ast}(t_{\rm ej}) + (t - t_{\rm ej})
v_{\rm exp}(t_{\rm ej}) \] the distance from the \cen\ of the star,
$R_{\ast}$ the stellar radius; $t_{\rm ej}$ stands for the time when a
certain layer was ejected, $v_{\rm exp}$ is the expansion velocity of the
wind at the moment of ejection.  It is assumed to be constant thereafter
and hydrodynamical effects are neglected (see also \sct~\ref{exp:vel}).  In
particular, the post-AGB wind has a higher velocity than the AGB shell and
will eventually overtake it.  That part of the post-AGB wind which has done
so contains little mass and is simply discarded.

\subsection{The AGB circumstellar shell}

The AGB shell is the principal contributor to the \iras\ fluxes, and it is
necessary to have a good description of the stellar temperature as function of
the time during the AGB, in order to compute the mass loss history
using \req{bowen}.
The \mete\ relation discussed in \sct~\ref{trans:tim}
starts at temperatures roughly between 3000~K and 4000~K (the relations we
received from Bl\"ocker extend to lower temperatures than given in his paper).
A description of the AGB evolution for lower temperatures is lacking.
Unfortunately, the evolution of \teff\ and $L$ on the AGB is not shown in the
Bl\"ocker papers.  We therefore simply extrapolate the \mete\ relation
logarithmically to lower temperatures.  The `AGB' star then evolves according
to the extended relation.  The \mete\ relation rises rather steeply at the low
temperature end so that only a limited amount of extrapolation is necessary.
In this way we find reasonable start temperatures for the AGB.  Since our
method implicitly assumes that the luminosity remains constant during the AGB,
the Bowen mass loss rates decrease with temperature.  We investigate two
different cases of AGB mass loss in the extrapolated part of the \mete\
relation: the normal Bowen mass loss and a constant mass loss held at the value
of the Bowen mass loss rate at the first point of the \mete\ relation as given
by Bl\"ocker.
These choices do not have implications for the post-AGB evolution.

\subsection{Expansion velocities}
\label{exp:vel}

The observed mean expansion velocity of AGB winds is 15~\kms\ (Olofsson\q\
1993), and this will be used as the typical AGB outflow velocity.  During
the post-AGB phase the situation is different.  Slijkhuis \& Groenewegen
(1992) assumed that the post-AGB wind has the same outflow velocity
(15~\kms) as the AGB wind.  The escape velocity (and hence also the outflow
velocity) increases with temperature however.  The increasing radiation
pressure from the hotter star on the less dense post-AGB wind accelerates
the dust and will decrease the densities in the post-AGB wind.  The escape
velocity for a 10\,000~K star is already of the order of 100~\kms\ to
150~\kms.  In addition, Szczerba \& Marten (1993) found that dust was
accelerated to 150~\kms\ during the post-AGB phase.  We therefore assume an
expansion velocity of 150~\kms\ in the post-AGB phase.

One should realize that the scenario of a fast wind that follows a slow
wind results in a collision between the two winds (as shown by e.g.
Mellema\q\ 1993 and Frank et al.\q\ 1993).  When inspecting the plots of
Mellema (1993) we find that the effects of the colliding winds only start
to become significant when the radiation driven wind ($\dm_{\rm CPN}$) with
velocities in excess of thousands of kilometers per second has developed.
The effect of colliding winds will be neglected in the further calculations
in this paper.

\section{The model runs}
\label{mod:runs}

\begin{table}
\caption[The calculated runs for the 0.605~\zm\ track]
{The calculated runs for the 0.605~\zm\ track, with an initial
envelope mass of 2~\zm, and an AGB expansion velocity of 15~\kms.}
\label{runs}
\begin{tabular}{crrrc}
\hline
Run & $v_{\rm PAGB}$ & \pta & \ptb &Constant \\
    & (\kms)         & (d)  & (d)  &mass loss   \\
\hline
1&    15  &    100 & 50  & \phantom{yes}\llap{no}    \\
2&    15  &    100 & 50  & yes   \\
3&   150  &    100 & 50  & yes   \\
4&   150  &    100 & 50  & \phantom{yes}\llap{no}    \\
5&   150  &    125 & 75  & \phantom{yes}\llap{no}    \\
6&  1500  &    100 & 50  & \phantom{yes}\llap{no}    \\
7&   150  &    125 & 75  & yes   \\
\hline
\end{tabular}
\end{table}

In this section we will investigate the spectral evolution of a post-AGB
star by varying certain parameters that influence the mass loss rate, wind
velocity and the evolutionary speed of the central star.  We restrict
ourselves to the 0.605~\zm\ track and the emphasis will be on the infrared
properties of the dust shell around the star. We calculated a total of
seven runs.  Every individual run consists of four different series of
models.  These individual models represent post-AGB shells with carbon-rich
or oxygen-rich dust, both with and without dust formation in the post-AGB
wind.  We assumed the total mass of the AGB and post-AGB shell to be 2~\zm.
With the extrapolated \mete\ relation this implies a start temperature of
the AGB of 2543~K.  The different run parameters are outlined in
\tbl~\ref{runs}.

The main differences between the runs are threefold. Firstly, as stated above,
the post-AGB wind velocity is subject to uncertainty. In order to assess the
influence of this velocity we used three different values: 15~\kms, 150~\kms\
and 1500~\kms.  Secondly, the differences between a decreasing AGB mass loss
rate and a constant mass loss rate are investigated. In the second case the
constant mass loss is kept at the value of the Bowen mass loss rate at the
first point of the \mete\ relation as given by Bl\"ocker. This value for the
mass loss of \dm = 1.99\xt{-4}~\zmy\ will then be adopted throughout the
extrapolated part of the \mete\ relation, i.e.\ for temperatures between 2543~K
and 3743~K. Approximately 85~\% of the total shell mass is ejected in this
phase. Thirdly, the definition of the end of the AGB is adjusted. As an
alternative the pulsation periods that define the start of the transition wind and the
start of the post-AGB phase are set to 125~d and 75~d respectively, which
corresponds to a transition phase occurring between 4723~K and 5418~K.  For
these runs the start of the transition period (defined by \pta) will be reached earlier,
however the transition period itself will take much longer due to the fact that the
\mete\ relation is much steeper now in the transition region, making the
evolutionary speed much slower. This gives the paradoxical result that an
earlier start of the transition phase gives rise to a later start of the post-AGB phase.
Also the smaller mass loss rates result in a slower evolution between 5418~K to
6042~K (the start of the post-AGB phase in the (\pta,\ptb) = (100d,50d) runs).
After the star has reached \teff~= 6042~K, the post-AGB mass loss rates are the
same, and consequently the evolutionary speed in terms of kelvin per year is
identical.

The models are calculated at the temperatures listed in \tbl~\ref{runtime}.
The second and third entries in this table reflect the time that has passed
since the end of the transition wind for the (\pta,\ptb) = (100d,50d) and
(\pta,\ptb) = (125d,75d) case respectively.

\begin{table}
\caption[The ages of the different models since the end of the AGB
transition phase]
{The ages of the different models since the end of the AGB transition
phase.  The age of the AGB shell, in the case of the (\pta,\ptb) =
(100d,50d) models (= normal end), is 6465~yr in the case of a decreasing
mass loss and 10\,326~yr in the case of a constant
mass loss.  For the (\pta,\ptb) =
(125d,75d) models (=early end), these ages are 6741~yr and 10\,603~yr
respectively.}
\label{runtime}
\begin{tabular}{rrr@{\hspace{1.5mm}}r|rrr}
\hline
\teff\ & {\small \hspace{-1mm}t(normal)} & {\small t(early)} & &
\teff\ & {\small \hspace{-1mm}t(normal)} & {\small t(early)} \\
(K)   &   (yr)  &  (yr)   & & (K)   &   (yr)  &  (yr)  \\
\hline
 5\,450 &         &  553    & &  9\,000 &   904\m & 5510    \\
 5\,500 &         & 1308    & &  9\,500 &   926\m & 5532    \\
 5\,550 &         & 1950    & & 10\,000 &   948   & 5553    \\
 5\,600 &         & 2492    & & 11\,000 &   991\m & 5597    \\
 5\,700 &         & 3328    & & 12\,000 &  1036   & 5642    \\
 5\,800 &         & 3904    & & 13\,000 &  1082\m & 5688    \\
 6\,000 &         & 4528    & & 14\,000 &  1131   & 5736    \\
 6\,050 &    13   & 4619\mm & & 15\,000 &  1180\m & 5786    \\
 6\,100 &    89   & 4695\mm & & 16\,000 &  1232   & 5838    \\
 6\,150 &   154   & 4760\mm & & 17\,000 &  1286\m & 5892    \\
 6\,200 &   211   & 4817\mm & & 18\,000 &  1342   & 5947    \\
 6\,250 &   262   & 4867    & & 19\,000 &  1399\m & 6004    \\
 6\,500 &   457   & 5063    & & 20\,000 &  1457   & 6063    \\
 7\,000 &   715   & 5320    & & 25\,000 &  1768   & 6374    \\
 7\,500 &   823   & 5429    & & 30\,000 &  2093\m & 6699    \\
 8\,000 &   860   & 5466    & & 35\,000 &  2409   & 7014    \\
 8\,500 &   883\m & 5488    & &         &         &         \\
\hline
\end{tabular}\par
\vbox{\small\hsize=\columnwidth\parindent=0mm\raggedright%
$^{\dagger}$ These models are not shown in the graphite tracks in
\figs~\ref{run4} and \ref{hotdust}.\par
$^{\ddagger}$ These models are not shown in the tracks in
\fig~\ref{run5}.}
\end{table}

\begin{figure*}
\begin{center}
\mbox{\epsfxsize=0.75\textwidth\epsfbox[24 315 535 668]{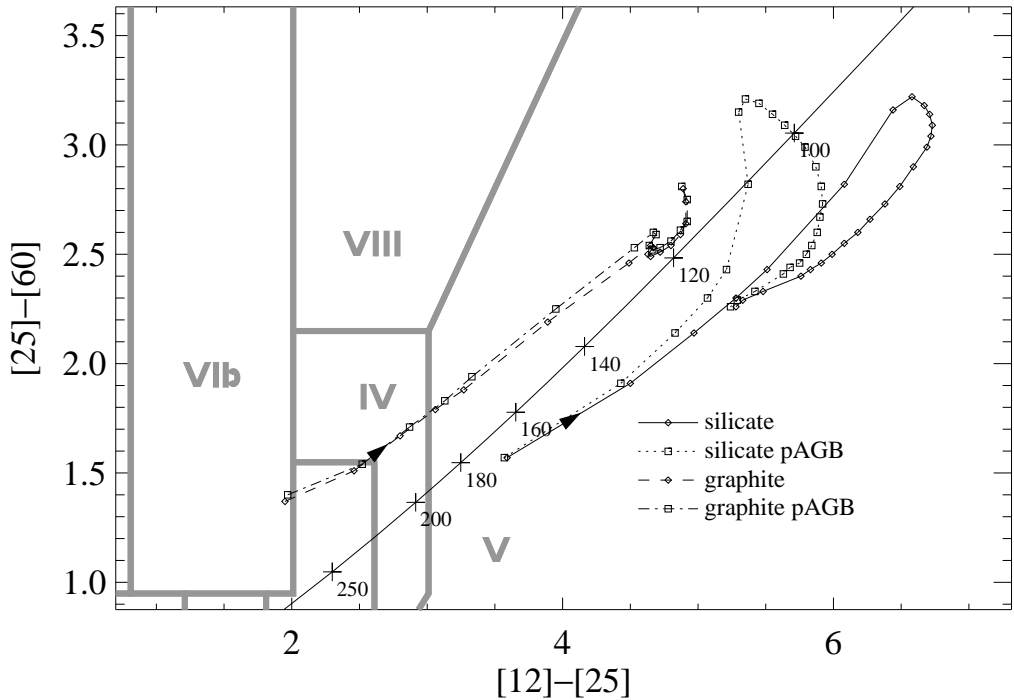}}
\caption[The \iras\ \cc\ diagram for run 4]
{\iras\ \cc\ diagram -- run 4. The arrows indicate the evolution of the
\iras\ \col\ with time.  The definitions for the \iras\ \cols\ are as
presented in Oudmaijer et al.\ (1992).  With these definitions the
Rayleigh-Jeans point, where objects hotter than approximately 2000~K are
found, is located at (0,0).  The boxes as defined by van der Veen \& Habing
(1988) are drawn in.  The straight line represents the blackbody
line. Selected temperatures are indicated. The first data point for all
tracks is the \teff\ = 6050~K model.}
\label{run4}
\end{center}
\end{figure*}

\begin{figure}
\begin{center}
\mbox{\epsfxsize=0.47\textwidth\epsfbox[20 145 493 718]{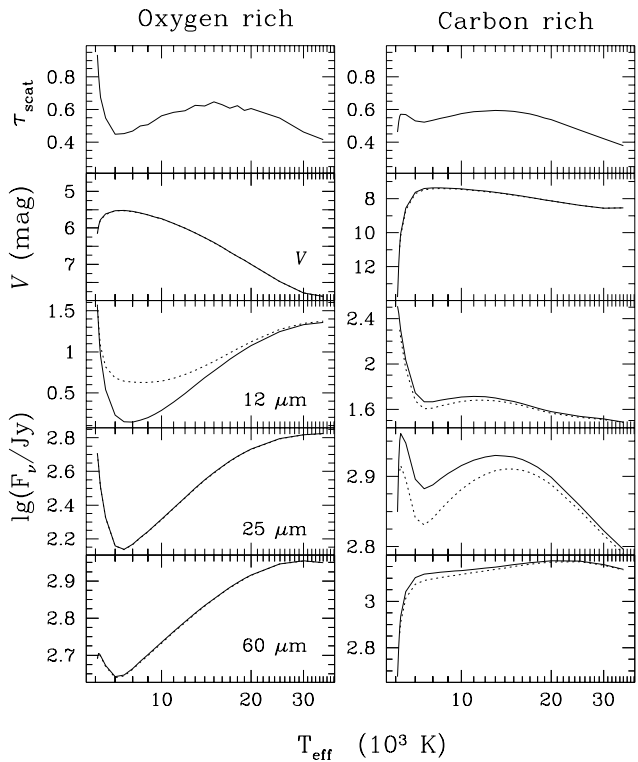}}
\caption[The \jv\ magnitude and the 12~\mic, 25~\mic\ and 60~\mic\
flux densities as a function of temperature]
{The \jv\ magnitude and the 12~\mic, 25~\mic\ and 60~\mic\ flux densities
as function of temperature during the post-AGB evolution for the 0.605~\zm\
run 4 models. The solid lines represent the models without dust formation
in the post-AGB outflow. The dashed lines indicate the models with dust
formation in the post-AGB wind. In the panels where no dashed lines are
visible, the values are indistinguishable. All curves start at \teff~=
6100~K, except the curves for graphite with post-AGB dust formation which
start at 6200~K.}
\label{run4flux}
\end{center}
\end{figure}

\begin{figure}
\begin{center}
\mbox{\epsfxsize=0.45\textwidth\epsfbox[30 315 535 668]{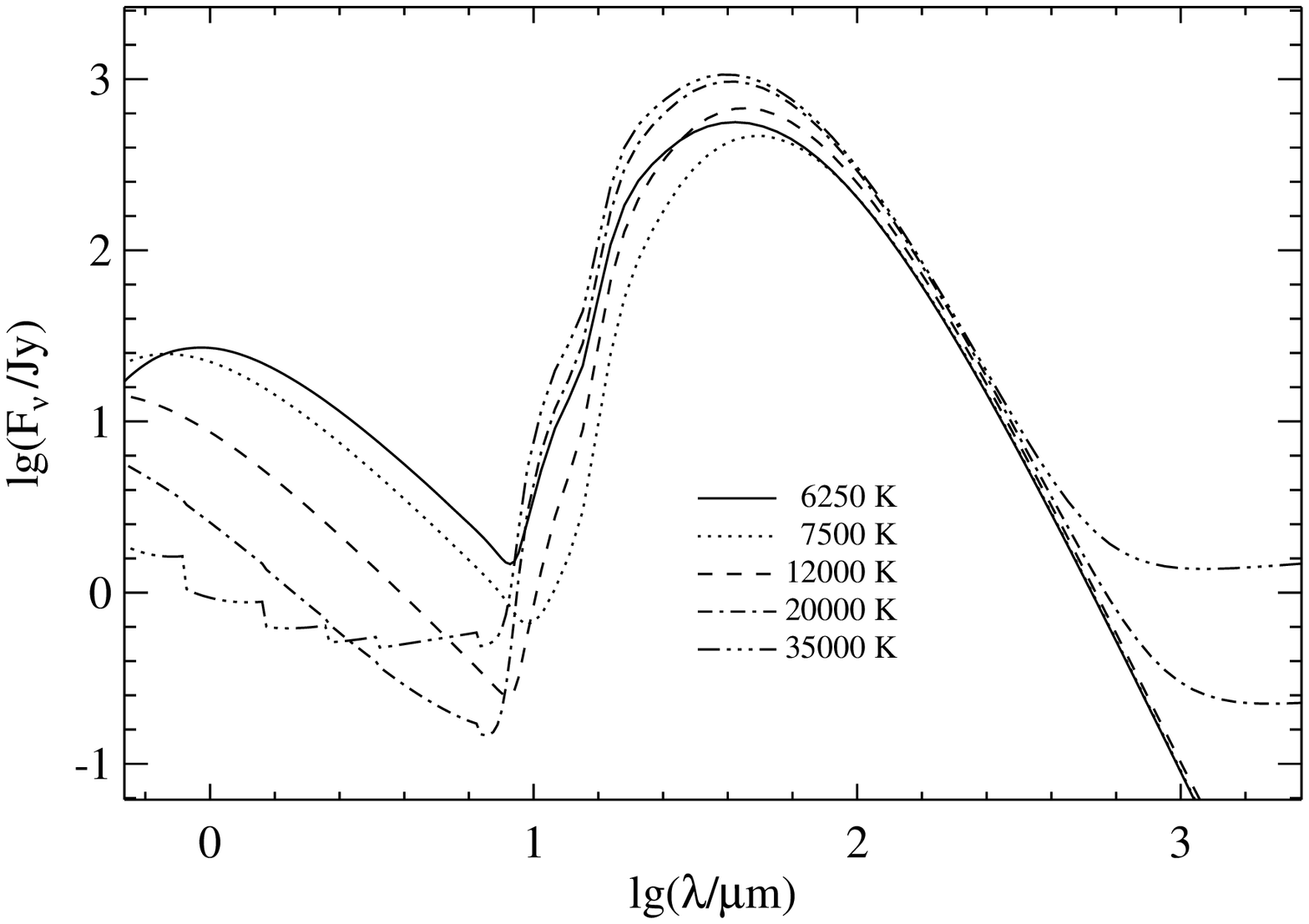}}
\vskip\floatsep
\mbox{\epsfxsize=0.45\textwidth\epsfbox[30 315 535 668]{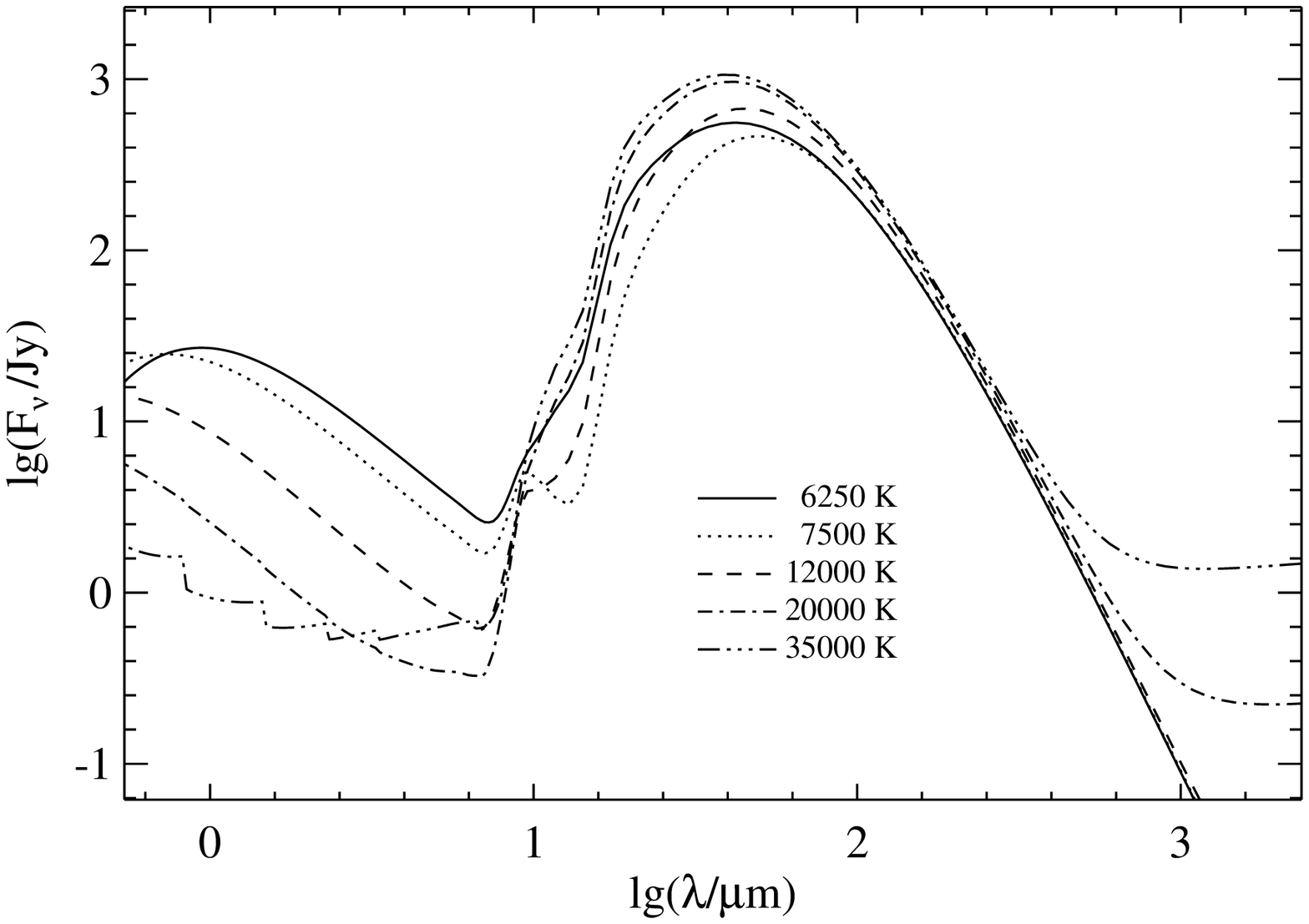}}
\caption[The spectral energy distribution at several stages during the
post-AGB evolution (run 4, silicates)]
{The spectral energy distribution at several stages during the post-AGB
evolution. The upper panel shows the SED for silicate dust without post-AGB
dust formation, the lower panel with dust formation, both for run 4.  The
flux densities are calibrated for an assumed distance of 1~kpc.  The
temperatures are indicated.  Note the presence of the silicate emission
feature at 10~\mic\ in the post-AGB dust models.}
\label{run4sedsil}
\end{center}
\end{figure}

\begin{figure}
\begin{center}
\mbox{\epsfxsize=0.45\textwidth\epsfbox[30 315 535 668]{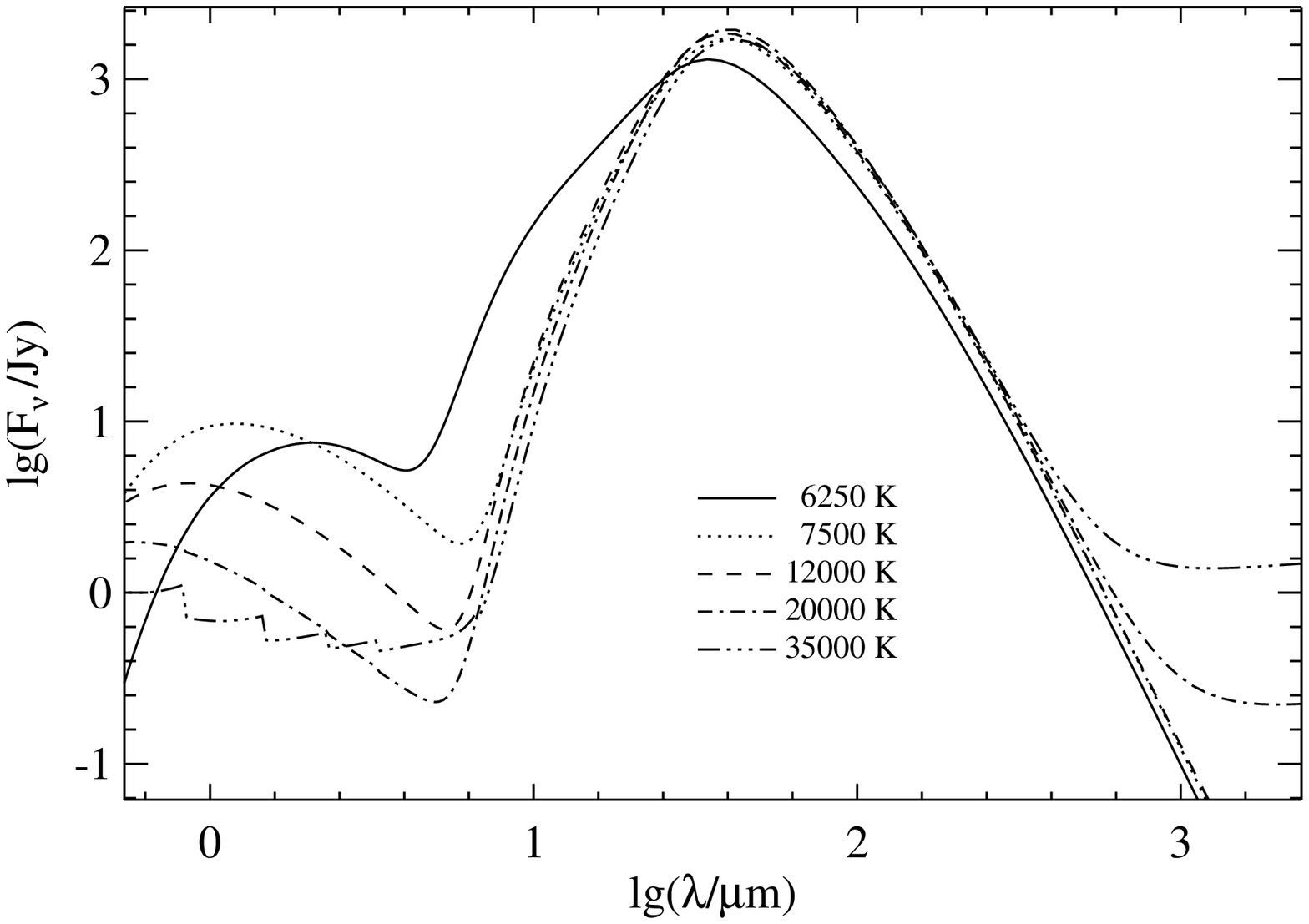}}
\vskip\floatsep
\mbox{\epsfxsize=0.45\textwidth\epsfbox[30 315 535 668]{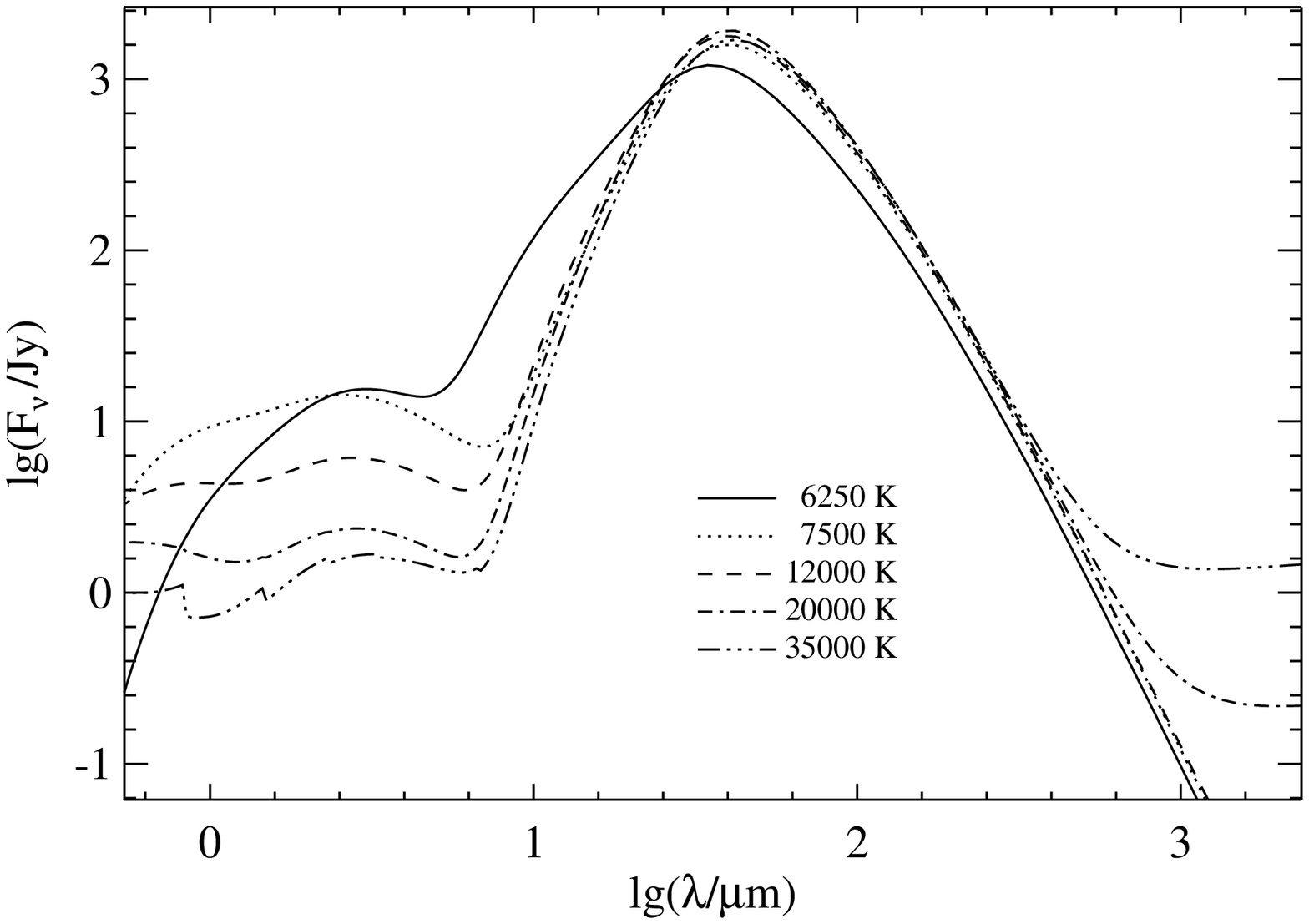}}
\caption[The spectral energy distribution at several stages during the
post-AGB evolution (run 4, graphite)]
{As the previous figure, but now for graphite with and without post-AGB
dust formation (lower panel, respectively upper panel).  Note the hot dust
that emits strongly in the near-infrared, but hardly contributes to the
12~\mic\ flux density.
}
\label{run4sedcar}
\end{center}
\end{figure}

We will start with the results of run 4, which we will use as a reference
since we consider the parameters for this run the most realistic.
Subsequently we discuss the main differences between this model run and the
other runs.
The main results of run 4 are visualized in \fig~\ref{run4}, where the
tracks in the \iras\ \cc\ diagram are plotted, in \fig~\ref{run4flux} where
certain fluxes are plotted, and in \figs~\ref{run4sedsil} and
\ref{run4sedcar} where a selection of spectral energy distributions is
presented. In these figures a distance of 1~kpc to the object is assumed.
We will discuss the results for the oxygen-rich and carbon-rich dust
separately.

For this run we have made an estimate for the
wavelength-averaged scattering optical depth
(see also the discussion in \sct~\ref{model}). For the models without
post-AGB dust formation the results are shown in \fig~\ref{run4flux}.
The results for the models with post-AGB dust formation are not shown, but they
are almost identical.
It can be seen that the scattering optical depth for the silicate and graphite
models are very similar, while on the other hand the absorption optical depth
is much higher for the graphite models when compared to the silicate models
(this can be judged by comparing the V magnitudes for both models).
This is caused by a combination of two effects.
Firstly, graphite is a very efficient absorber at optical wavelengths and a
relatively less efficient scatterer. The reverse is the case for silicates.
This tends to level out the differences. Secondly, in the averaging process the
optical depths are weighted by the output spectrum. This spectrum peaks much
more towards the red for the graphite models due to the higher internal
extinction.
The scattering optical depth is lower at longer wavelengths and this tends to
level out the differences even further. The latter effect also explains why
the scattering optical depth drops off so slowly, or even rises towards higher
temperatures. The peak of the energy distribution shifts towards the blue,
where the scattering efficiency is much higher, thus countering the effects
of the dilution of the circumstellar envelope.

Using these results we were able to obtain a worst case estimate for the amount
of energy that could have been missed in the dust emission due to the fact that
the scattering processes were not properly treated in our models. The amounts
for the models shown in \fig~\ref{run4flux} are typically 10~\% to 20~\% of
the total far-IR flux for the silicate models, and 5~\% to 10~\% for the
graphite models. The worst case estimates for the models not shown in
\fig~\ref{run4flux} are all below 40~\%.
These estimates reflect on the accuracy of the absolute fluxes and magnitudes
predicted by our models. However, they are expected to cancel in a first order
approximation when \cols\ are calculated. Therefore we decided to omit the
lowest temperature models whenever absolute fluxes are shown, but to include
them when \cols\ are shown.

Scattering processes will not only influence the
total amount of energy absorbed, but also the run of the dust temperature with
radius. This can influence the shape of the spectrum and hence also the \cols.
It is impossible to make an estimate for the magnitude of this effect and we
will assume it to be negligible.

\subsection{Silicate dust}

We will start with a description of the silicate tracks in the \iras\ \cc\
diagram (\fig~\ref{run4}). At first the \colii\ and \coliii\ \cols\ both
increase.  At some point the \coliii\ \col\ remains constant, while \colii\
still increases.  This is followed by a decrease in both the \colii\ and
\coliii\ \cols. Hence the track makes a clockwise loop in the \cc\ diagram.

The first part of the track represents the cooling of the dust shell due to
its expansion.  From \tbl~\ref{runtime} one sees the inner radius of the
shell has increased by a factor of 4 between \teff\ = 6200~K and \teff\ =
8000~K, which leads to a cooling and a decrease of the \iras\ fluxes
(\fig~\ref{run4flux}).  The following evolution in the \iras\ \cc\ diagram
is rather counter-intuitive.  Normally one would expect that the shell
would make the familiar counter-clockwise loop in the \cc\ diagram as found
by Loup (1991), Volk \& Kwok (1989) and Slijkhuis \& Groenewegen (1992).
In such a loop, the shell continues to cool, until the photospheric
radiation begins to dominate the emission, first at the 12~\mic\ band.  At
this time, the 25~\mic\ flux and the \colii\ \col\ decrease strongly due to
the expansion and cooling of the shell.  The \coliii\ \col\ will continue
to increase slowly.  Later, as the star starts to dominate at 25~\mic, the
\colii\ \col\ will remain constant and the \coliii\ \col\ will start to
decrease.  When the star is the dominant contributor in all three \iras\
bands, the loop will end in the Rayleigh-Jeans point.  However, the above
authors assumed a constant temperature of the central star, while the
effects of an evolving star on the \iras\ \cols\ can not be neglected; in
the 0.605~\zm\ track the star rapidly evolves toward higher effective
temperatures.  For example, the kinematic age of the shell, and thus its
outer (or inner) radius, has increased by only 25~\% between \teff\ =
8000~K and \teff\ = 12\,000~K.  One can regard the circumstellar shell as
essentially stationary around the evolving star.  An evolving star embedded
in a stationary shell results in a heating of the dust and an increase of
the \iras\ fluxes (\fig~\ref{run4flux}). The effect of re-heating
of the circumstellar envelope was already found by Marten, Szczerba \&
Bl\"ocker (1993) in their calculations. The re-heating continues even
beyond the last data point, hence in the \cc\ diagram the track keeps
evolving to higher \col\ temperatures in both \cols.  Around the turning
point the 12~\mic\ flux is partly due to the photosphere.  The photospheric
flux decreases for higher temperatures so that the 12~\mic\ flux reacts
later to the rising dust flux than the 25~\mic\ and 60~\mic\ flux.
Therefore the \colii\ \col\ starts decreasing later than the \coliii\ \col.
The result is a clockwise loop in the \cc\ diagram.

One should note the \beh\ of the \iras\ flux densities in
\fig~\ref{run4flux}.  The flux densities reach a minimum during the slow
evolution of the star.  The minimum is followed by gradual increase in the
total infrared output due to the heating of the shell resulting from the
increasing absorbing efficiency of the dust (see also the discussion in
\sct~\ref{carbon:dust}).  A model star like this (i.e.\ with silicate dust)
would have a larger chance of being detected in an infrared survey like
\iras\ when it has evolved to higher temperatures.  This fact, combined
with the predicted distribution over spectral type implies that hot
post-AGB stars, or young PN should be expected to be more abundant in
samples of evolved stars with infrared excess.

\subsubsection{Dust formation in the post-AGB wind}

The addition of dust formation in the post-AGB wind changes the spectral
energy distribution.  From \fig~\ref{run4flux} it is visible that the
25~\mic\ and 60~\mic\ flux densities have the same values as in the
AGB-only dust formation case.  The 12~\mic\ band is strongly affected by
the addition of the hot dust.  This is due to the presence of the 10~\mic\
silicate emission feature which causes the emissivity of the hot silicate
dust to peak strongly at 10~\mic.  The larger 12~\mic\ flux density is
immediately reflected in the \cc\ diagram.  It puts the starting point at a
slightly higher \col\ temperature than before.  As the feature increases in
strength, the track bends toward higher \colii\ \col\ temperatures.  Then,
when the star is evolving more rapidly than the shell expands, the cool AGB shell
is heating up, and the track moves down.  At the same time, the \colii\
\col\ temperature decreases, reflecting a weakening of the silicate feature
relative to the 25~\mic\ flux.  This effect is caused by the circumstances
under which dust in the post-AGB wind is formed.  As the star becomes
hotter, also the distance increases at which the equilibrium temperature of
the dust goes below 1000~K.  Therefore the dust formation will take place
at larger distances from the star, in a diluted local radiation field where
the dust density will be lower.  Consequently, the silicate feature weakens
and the track moves to lower \colii\ \col\ temperatures.  Eventually the
contribution of the 10~\mic\ feature will become negligible and the track
will move asymptotically to the track without post-AGB dust formation.

\subsubsection{Different post-AGB wind velocities} 

\begin{figure*}
\begin{center}
\vskip5mm
\mbox{\epsfxsize=0.75\textwidth\epsfbox[24 315 535 668]{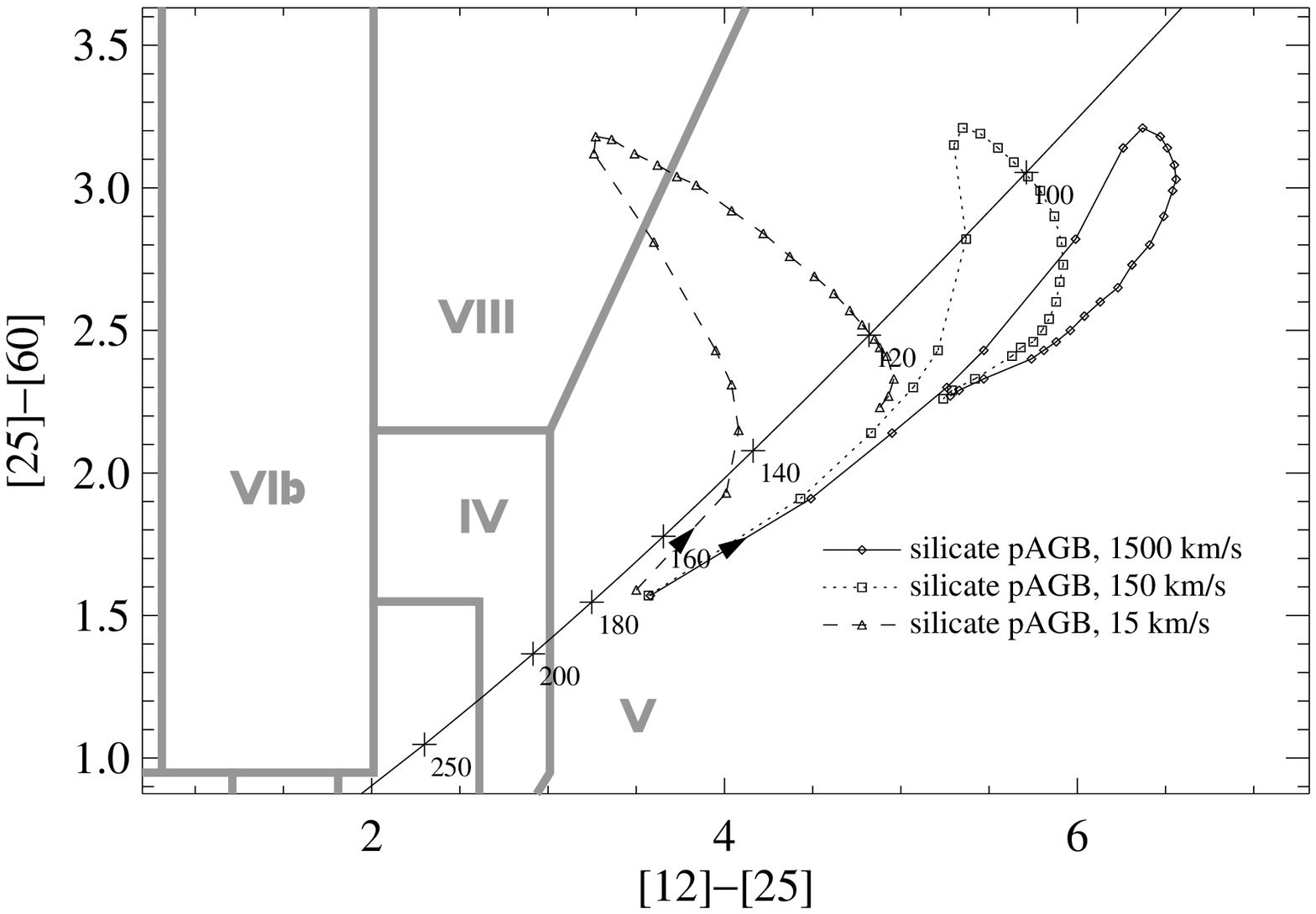}}
\vskip9mm
\mbox{\epsfxsize=0.75\textwidth\epsfbox[24 315 535 668]{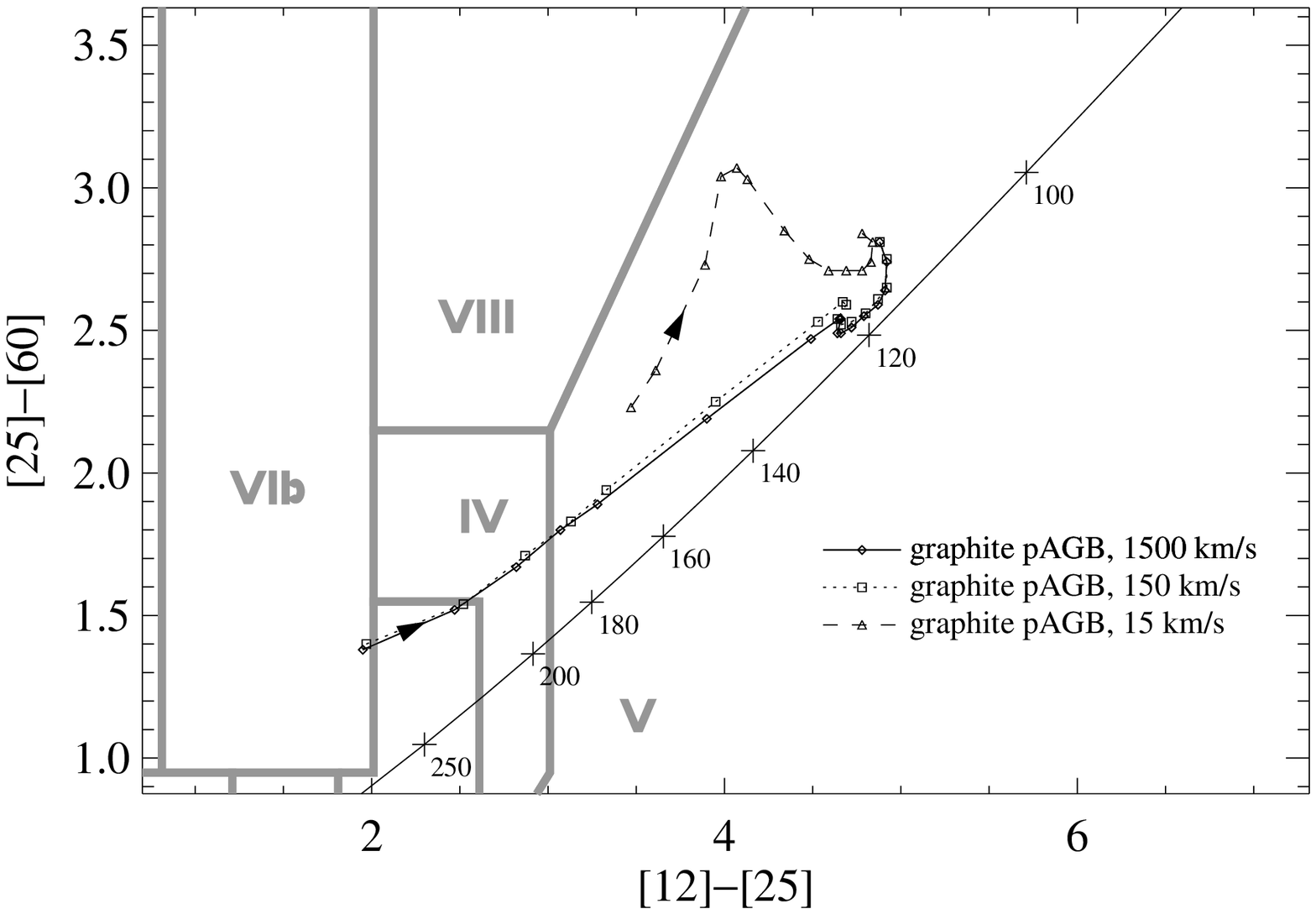}}
\vskip3mm
\caption[The \iras\ \cc\ diagram for different
velocities of the post-AGB wind]
{In these \iras\ \cc\ diagrams, the effect of different velocities of the
post-AGB wind with dust formation is presented.  The 15~\kms, 150~\kms\ and
1500~\kms\ models are calculated in runs 1, 4 and 6.  The upper diagram
shows the silicate models, the lower diagram the graphite models.  The
graphite 15~\kms\ track starts at 6200~K, all others at 6050~K.}
\label{hotdust}
\end{center}
\end{figure*}

In \fig~\ref{hotdust} the effect of a different post-AGB wind velocity is
shown for silicate dust in the upper panel.  The models have been
calculated for velocities of 15~\kms, 150~\kms\ and 1500~\kms.  The
1500~\kms\ tracks resemble the AGB-only dust formation models, because the
amount of dust in the post-AGB wind is not large enough to yield an
observable effect.  The low outflow velocity model shows all the effects
outlined for the 150~\kms\ models even stronger because the amount of hot
dust has increased.

\subsection{Carbon-rich dust}
\label{carbon:dust}

Let us now turn to the models with carbon-rich dust.  This type of dust is
a more efficient absorber than silicate dust as can be seen in
\fig~\ref{run4flux}. The visual magnitude is larger than for the oxygen
rich models, reflecting a larger circumstellar extinction in the
Johnson~\jv\ band.  One can also see that the infrared energy output is
larger.

The beginning of the track of the carbon-dust models in the \cc\ diagram
(\fig~\ref{run4}) is located at a higher \col\ temperature than for the
oxygen-rich models because more energy is absorbed by the dust.  At first,
the track moves to lower temperatures, and as the star begins to evolve
more rapidly, the curve makes a slight backward loop in the diagram due to the
heating of the shell.  Then, contrary to the silicate model, the shell
cools again.

This somewhat unexpected result can be explained by the properties of the
absorbing material.  It is instructive to investigate the effective cross
section $Q_0$ of the two grain species as a function of the effective
temperature of the central star spectrum (i.e.\ a blackbody).  $Q_0$ is
defined as

\begin{equation}
Q_0 = \frac
{\int_0^{\infty}B_{\nu}(T_{\rm eff})\alpha_{\nu}{\rm d}\nu}
{\int_0^{\infty}B_{\nu}(T_{\rm eff}){\rm d}\nu}
\label{dust:int}
\end{equation}

\begin{figure}
\begin{center}
\mbox{\epsfxsize=0.45\textwidth\epsfbox[30 312 535 668]{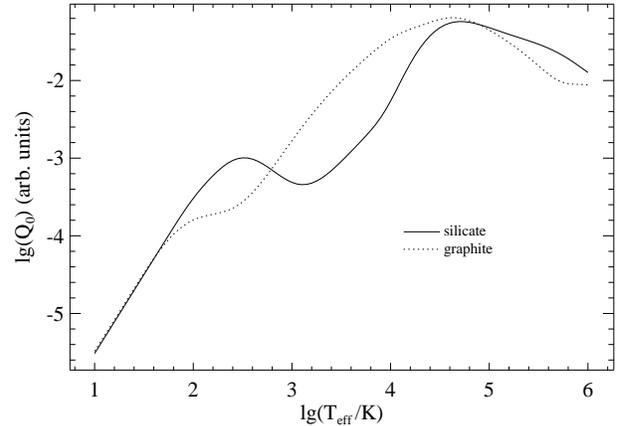}}
\caption[The total amount of energy absorbed by grains
as a function of temperature]
{The effective cross section for the grain species used in our study as a
function of the effective temperature of a blackbody.  The total luminosity
of the central source remains constant and the dust-to-gas mass ratio for
both grain species is equal.}
\label{dust:plot}
\end{center}
\end{figure}

With $B_{\nu}$ the blackbody intensity distribution and $\alpha_{\nu}$ the
absorption cross section of the grains.  The formula is chosen in such a
way that the total energy absorbed by the grains is proportional to
$L\times Q_0$. Note that $L$ is constant during the evolution. 
The absorption coefficients have been chosen
such that both grain species have the same dust-to-gas mass ratio.  The
resulting curves
are shown in \fig~\ref{dust:plot}

For temperatures between approximately 1000~K and 50\,000~K silicates
absorb less energy than graphite.  It can be a factor of 10 lower in the
temperature range for cool post-AGB stars.  This is caused by the fact that
silicates are inefficient absorbers at optical wavelengths.  From these
results we can expect that for temperatures between 10\,000~K and 50\,000~K
the total IR emission of silicates will rise much more drastically than for
graphite.  For temperatures above 50\,000~K the peak of the energy
distribution shifts into the EUV.  At these wavelengths silicates are more
efficient absorbers and we can see that for these temperatures the total
amount of energy absorbed by the silicates is higher than for graphite.

Thus, in the last data points of the track in the \iras\ \cc\ diagram, the
absorption efficiency of graphite does increase less strongly with rising
effective temperature.  In this case, the interplay between the evolving
star (now heating up the shell less rapidly) and the expanding shell
(giving rise to cooling) has as a result that the expanding shell becomes
the more dominant factor and consequently the shell cools.  In
\fig~\ref{run4flux} these two effects are illustrated by the decreasing
\iras\ fluxes.

\subsubsection{Dust formation in the post-AGB wind}

In contrast to oxygen-rich dust, graphite has no feature in the \iras\
12~\mic\ pass band.  Also the shape for the emissivity law of graphite is
such that the hot dust peaks at near-infrared wavelengths, contributing
negligibly to the 12~\mic\ band (\figs~\ref{run4flux} and
\ref{run4sedcar}).  Therefore one does not expect a large effect of the hot
dust on the \iras\ 12~\mic\ flux.  However, the 12~\mic, 25~\mic\ and
60~\mic\ flux densities of these models are lower than for the
corresponding AGB-only dust formation models (\fig~\ref{run4flux}).  This
stems from the fact that the hot dust absorbs a large fraction of the
stellar energy, yielding a cooler AGB shell.

\subsubsection{Different  post-AGB wind velocities} 

The effect of the post-AGB outflow velocity on the graphite models is
presented in the lower panel of \fig~\ref{hotdust}.  The 1500~\kms\ track
behaves almost the same as the 150~\kms\ track and the track without
post-AGB dust.  The 15~\kms\ track however is different, first it moves
upward, then bends downward and finally makes the counter-clockwise loop.
The initial increase in the \coliii\ \col\ is caused by the fact that the
large amount of hot dust now shields the AGB shell more efficiently, making
the AGB shell even cooler than previously.  As the star becomes hotter, the
dust condensation radius moves rapidly outward, making shielding less
efficient.  The AGB shell now obtains more energy from the central star and
heats up, resulting in higher \coliii\ temperatures.  The tracks evolve
asymptotically towards each other and finally the counter-clockwise loop
sets in, as was already explained.

\subsection{Constant mass loss}

\begin{figure*}
\begin{center}
\mbox{\epsfxsize=0.75\textwidth\epsfbox[24 315 535 668]{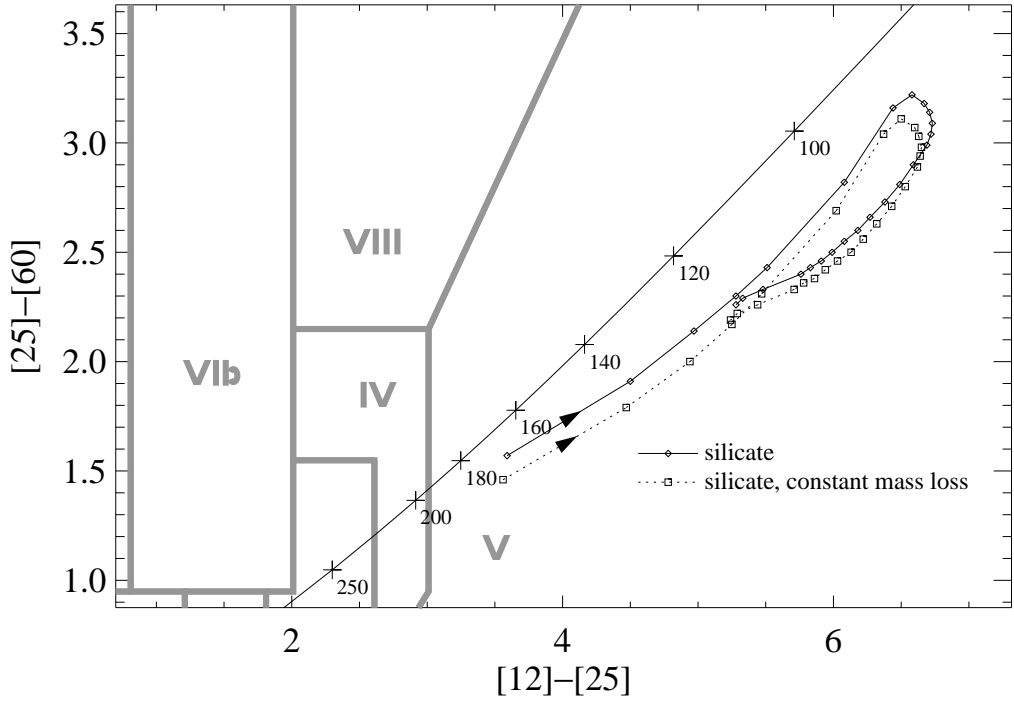}}
\caption[The \iras\ \cc\ diagram, comparing decreasing and constant
mass loss]
{\iras\ \cc\ diagram in which the difference between the decreasing and
constant mass loss (runs 4 and 3) are shown for the silicate tracks. The
first data point for both tracks is the \teff\ = 6050~K model.}
\label{cagb}
\end{center}
\end{figure*}

In \fig~\ref{cagb} the silicate tracks for run 4 (decreasing AGB mass loss)
and run 3 (constant mass loss) in the \iras\ \cc\ diagram are shown. Both
the \colii\ and the \coliii\ \col\ temperature of the constant mass loss
track are slightly higher than for the decreasing mass loss track.

The constant mass loss creates relatively more cool dust further from the
star.  This cooler dust emits less radiation and therefore the hotter dust
at the inner parts of the AGB shell (where the mass loss rates for both
tracks are the same) will dominate the spectrum more. This gives the result
that the integrated spectrum of all dust appears hotter. This effect is
strongest at 60~\mic, less at 25~\mic\ and absent at 12~\mic.  As a whole,
the effect of different AGB mass loss rates is small.  The effect on the
graphite tracks is basically the same and is not shown.

\subsection{Different end of the AGB}

For all the models that have been discussed so far, it was assumed that the
transition mass loss starts at \pta\ = 100~d, and that the Reimers mass loss starts
at \ptb\ = 50~d. However, the evolutionary \beh\ of the calculated models
strongly depends on the exact moment of `superwind' cessation (Szczerba\q\
1993). In this section, we will present an investigation into the effect of
changing this unknown parameter.

As stated before, the evolution of the central star in the (\pta,\ptb) =
(100d,50d) and the (\pta,\ptb) = (125d,75d) models is identical after
\teff\ = 6042~K. However, the density structure of the circumstellar shell
is entirely different when the star has reached this temperature.  Whereas
the (\pta,\ptb)~= (100d,50d) model then barely has a detached shell, the
(\pta,\ptb)~= (125d,75d) model already has a shell with a kinematic age in
excess of 4500~yr, which has much cooler dust.

\begin{figure*}
\begin{center}
\mbox{\epsfxsize=0.75\textwidth\epsfbox[24 315 535 668]{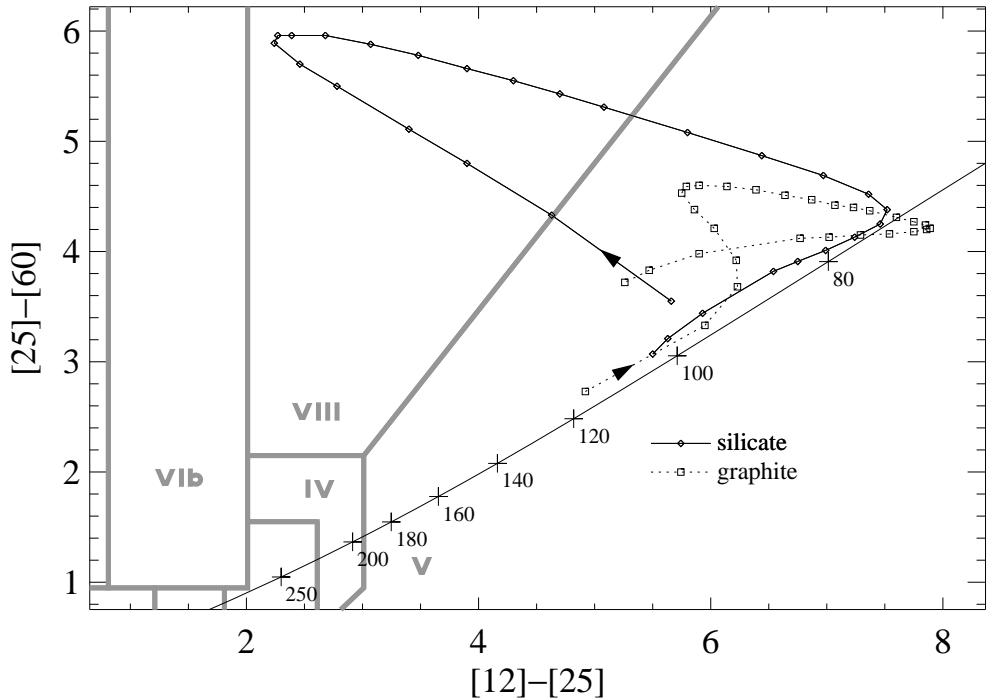}}
\caption[The \iras\ \cc\ diagram for run 5]
{\iras\ \cc\ diagram for the models without post-AGB dust formation in run
5, the case with an earlier start of the transition phase than in run 4. The first data
point for both tracks is the \teff\ = 5450~K model.}
\label{run5}
\end{center}
\end{figure*}

\begin{figure}
\begin{center}
\mbox{\epsfxsize=0.45\textwidth\epsfbox[30 315 535 668]{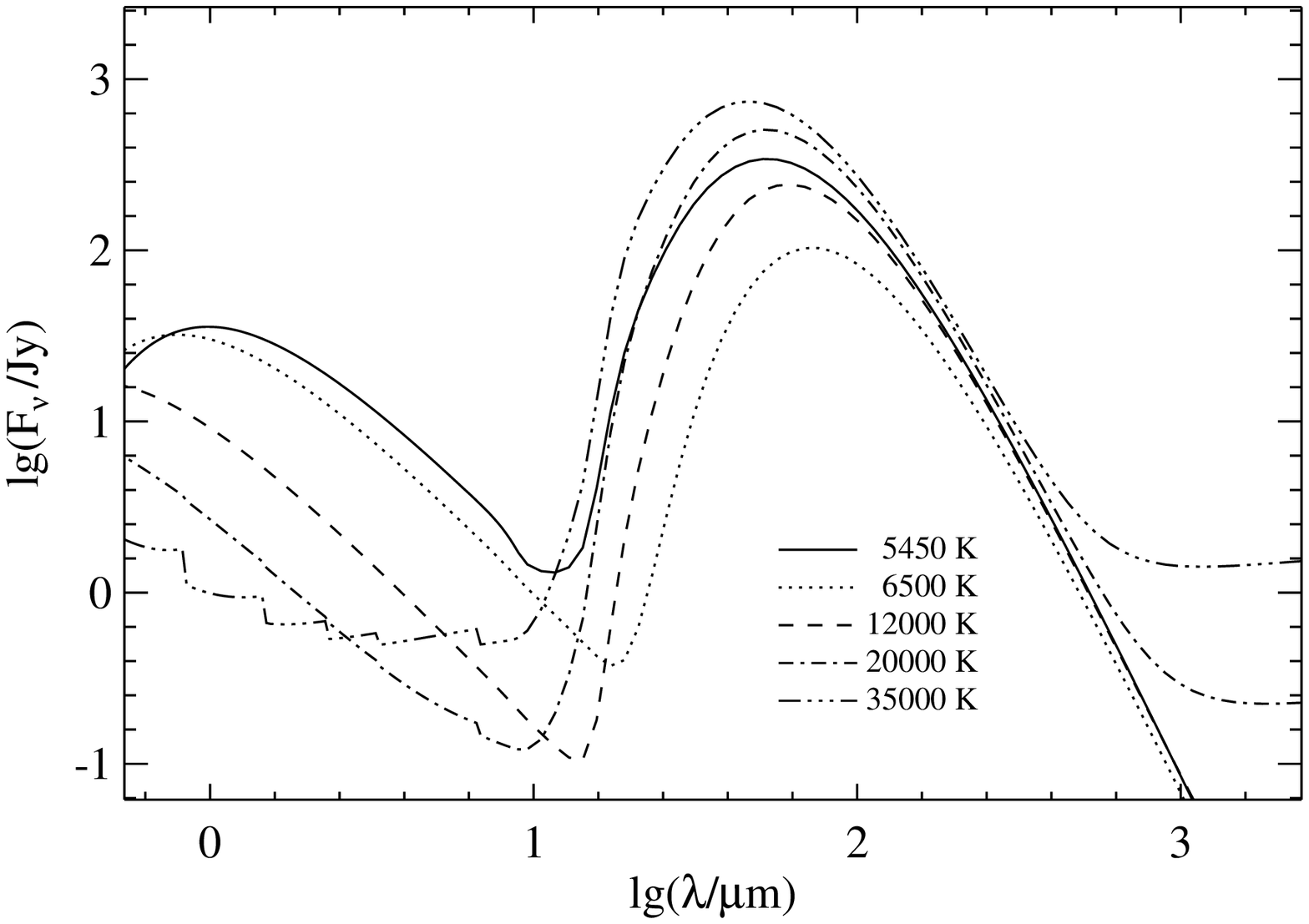}}
\vskip\floatsep
\mbox{\epsfxsize=0.45\textwidth\epsfbox[30 315 535 668]{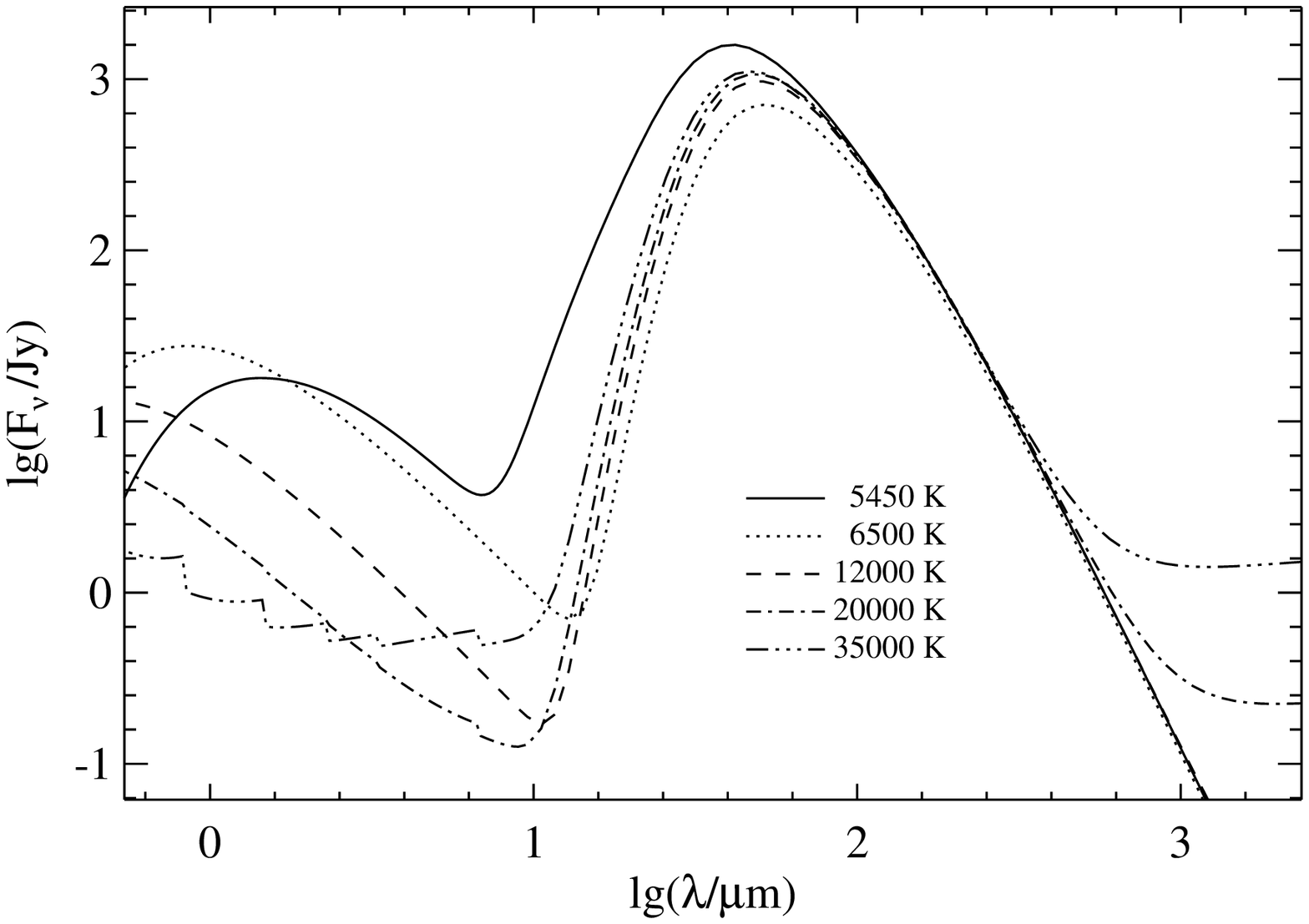}}
\caption[The spectral energy distribution at several stages during the
post-AGB evolution (run 5)]
{The spectral energy distribution at several stages during the post-AGB
evolution for an earlier start of the transition phase (run 5).  The upper panel
shows the SED for silicate dust without post-AGB dust formation, the lower
panel for graphite without post-AGB dust formation.  The flux densities are
calibrated for an assumed distance of 1~kpc.  The temperatures are
indicated.}
\label{run5sed}
\end{center}
\end{figure}

We will now look at the results for this run in the \cc\ diagram. It
appears that the alternative end of the AGB entirely changes the path in
the \cc\ diagram (\fig~\ref{run5}).  A selection of spectral energy
distributions is presented in \fig~\ref{run5sed}.

From the start, the silicate dust track moves strongly to the upper left,
making a turn to the lower right, and finally a bend to the left.  Since
the star evolves slowly (it takes approximately 750~yr to go from 5450~K to
5500~K), the circumstellar shell cools rapidly, and only the stellar
photosphere is visible at 12~\mic\ (\fig~\ref{run5sed}).  This is reflected
in an increasing \colii\ \col\ temperature: the photospheric 12~\mic\ flux
remains approximately the same, while the 25~\mic\ flux and the \coliii\
temperature decrease due to the cooling of the shell. Subsequently the star
starts to evolve more rapidly than the shell expands, and the \colii\ \col\
temperature decreases.  This is due to the fact that the shell now heats
up, increasing the 25~\mic\ flux and decreasing the \coliii\ \col, contrary
to the 12~\mic\ band which is still dominated by the decreasing
photospheric emission of the evolving star.  The shell continues to heat
up, and the dust emission at 12~\mic\ overtakes the decreasing photospheric
flux density.  The curve bends to higher \colii\ temperatures.  The
contribution of nebular bound-free emission at 12~\mic\ for the hottest
models increases the \colii\ \col\ temperature even more, and the track
makes the, now familiar, clockwise loop.

All effects outlined above appear less strong for the carbon-rich model.
It initially starts on a cooling curve, and later bends to the clockwise
loop.  This is explained by the fact that the dust emission from the carbon
grains still is present at 12~\mic\ in the first points, so that the shell
cools in all \cols.  For the remainder of the evolution there always
remains some dust emission in the 12~\mic\ band, making the loop to the
left less pronounced.

\subsection{A near-IR \cc\ diagram}

\begin{figure*}
\begin{center}
\mbox{\epsfxsize=0.75\textwidth\epsfbox[30 315 535 668]{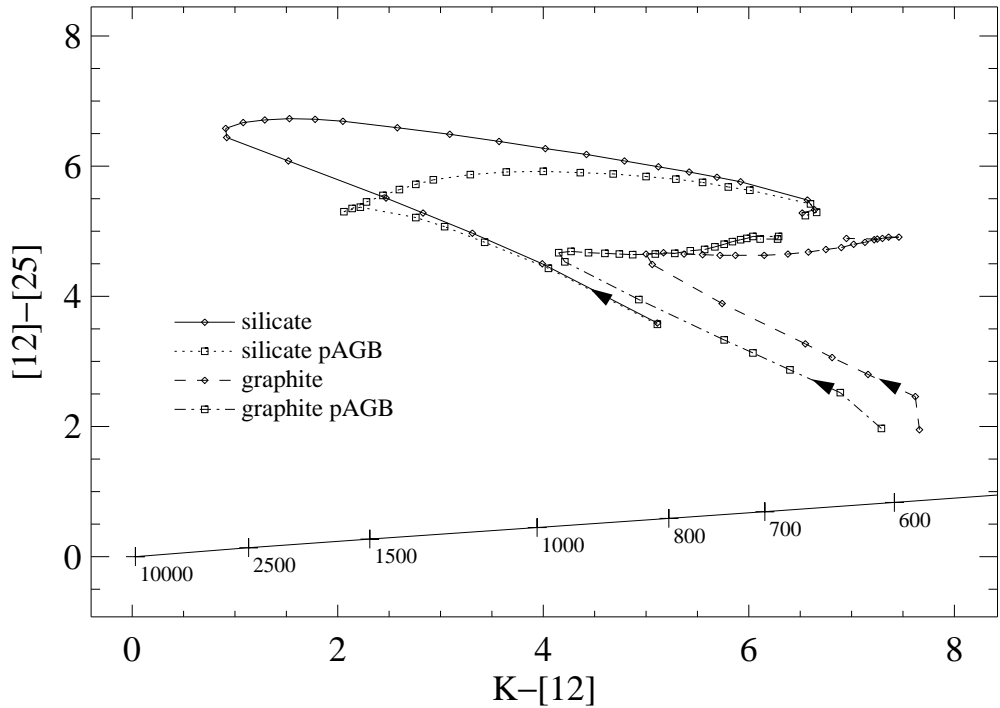}}
\caption[The near-IR \cc\ diagram (run 4).]
{The near-IR \cc\ diagram at several stages during the post-AGB evolution
for run 4. For reference a blackbody curve is also included in the
diagram.}
\label{run4ccalt}
\end{center}
\end{figure*}

In this section we present the evolution of a post-AGB star in a \cc\
diagram which uses \cols\ at shorter wavelengths: the \coli\ vs. \colii\
diagram.  This diagram gives a different view of the evolution since in all
models the Johnson \jk\ band will be dominated either by the stellar
continuum or the bound-free emission of the ionized part of the
nebula. Only in the models containing hot graphite grains, part of the flux
in this band will originate from the grains.  So, contrary to the \iras\
\cc\ diagram which we presented earlier, this diagram contains information
both on the central star and the dust.  The results for run 4 are shown in
\fig~\ref{run4ccalt}, giving both the tracks using silicate and graphite
grains and both the tracks assuming AGB-only and post-AGB dust formation.

The first thing we notice is that all four tracks, at least in a
qualitative sense, look quite similar. This suggests that this \cc\ diagram
is less sensitive to particulars of the grain emission and thus that the
information on the evolution of the central star and the nebula is less
`contaminated' when compared with the \iras\ \cc\ diagram.  We have only
investigated this for the 3~\zm\ track and it is not yet clear if this
observation is valid in a more general context.  In the past years the
\iras\ \cc\ diagram has proven to be a very useful tool for studying
post-AGB evolution. Our knowledge has increased considerably since its
introduction. However, when we try to understand the details of this
evolution better, the information from the \iras\ \cc\ diagram
becomes more and more confusing. The \coli\ vs. \colii\ diagram may be
a valuable additional tool for the study of post-AGB evolution.

We will only discuss the evolution of the \coli\ \col\ in
\fig~\ref{run4ccalt} since the evolution of the \colii\ \col\ already has
been discussed earlier.  At first both the flux in the \jk\ band and the
12~\mic\ band are decreasing.  However the 12~\mic\ flux decreases more rapidly
and therefore \coli\ evolves towards hotter \cols.  This can be understood
if we realize that the central star heats up relatively slowly while the
circumstellar shell expands relatively rapidly. At a temperature of around
8000~K the central star evolution will speed up considerably, reversing the
preceding argument. The flux in the \jk\ band continues to drop,
while the flux in the 12~\mic\ band increases now, resulting in
ever cooler \coli\ \cols. This evolution continues until the central star
starts to ionize a considerable part of the circumstellar shell and the
\jk\ band flux will start to rise again due to nebular bound-free
emission. This rise is so rapid that it reverses the evolution of the
\coli\ \col.

The tracks for silicate and graphite grains are qualitatively similar, but
are offset with respect to each other. This is mainly due to the different
absorption efficiency of silicate and graphite grains (see
\sct~\ref{carbon:dust}).  The difference between the silicate track with
and without post-AGB dust formation can be understood solely from the
presence or absence of the 10~\mic\ emission feature. The difference
between the graphite tracks with and without post-AGB dust formation 
stems from the fact that the hot dust in the post-AGB section of the
wind mainly radiates around 3~\mic\ and thus contributes to the \jk\ band
flux.

\section{Discussion and conclusions}

In this paper we have presented a new model to calculate the spectral
evolution of a hydrogen burning post-AGB star.
The main new ingredient of this model is the
possibility of extracting timescales of the post-AGB central star evolution
from the most recent evolutionary calculations.  Hence, contrary to
previous studies, it is possible now to investigate other AGB and/or
post-AGB mass loss rates and different prescriptions for the start of the
post-AGB phase.  The use of a photo-ionization code in which a dust code
is built in, gives us the opportunity to study the evolution of the
infrared emission of the circumstellar material.

We have performed a parameter study on a typical post-AGB star with a core
mass of 0.605~\zm\ taken from Bl\"ocker (1995b).  By varying the parameters
that govern the mass loss and the evolutionary timescales of the post-AGB
phase by a reasonable amount we find that:
\begin{enumerate}
\renewcommand{\theenumi}{(\arabic{enumi})}
\item
The influence of the evolving star on the infrared \col\ evolution can not
be neglected. In particular, the phase wherein the circumstellar shell can
be considered as a stationary shell around a star that rapidly increases in
temperature results in clockwise loops in the \iras\ \cc\ diagram. This was
not observed in many previous studies in which the temperature of the
central star was taken to be constant.  Instead, in these studies the
evolutionary path followed a counter-clockwise loop in the \cc\ diagram
because of the inevitable cooling of the shell. \\ Only Szczerba \& Marten
(1993) who used a Bl\"ocker track obtained a roughly similar result with
their dust radiative transfer code.  Volk (1992) found a slight deviation
from the counter-clockwise loop in the \cc\ diagram.  He used the coarse
grid of evolutionary timescales from the 0.644~\zm\ Sch\"onberner (1983)
track.  The output of the photo-ionization code \cld\ was used as input for
a dust radiative transfer model.  It is not clear to what extent that
choice influenced the path followed in the \iras\ \cc\ diagram. \\ The
crucial influence of the evolution of the central star warrants further
parameter studies with other stellar models, where the increase in
temperature as a function of time will be different. It is expected that
the heating of the shell will be more important for more rapidly evolving (i.e.
larger core mass) stars.
\item
The tracks that are followed in the \iras\ \cc\ diagram are very sensitive
to the adopted dust opacity law and solid state features.  Two main
differences between the silicate and graphite models that were studied are
evident.  Firstly the \iras\ \cols\ of the silicate models are very
sensitive to newly synthesized dust in the post-AGB wind because of the
silicate 10~\mic\ feature that contributes significantly to the 12~\mic\
flux density.  In contrast, hot graphite dust emits mainly shortward of the
12~\mic\ pass band and thus post-AGB dust formation has less influence in
this case.  Secondly, the dependency of the absorption efficiency on the
central star temperature is much stronger for silicates than for graphite
in the temperature regime studied here. Therefore silicates react much
stronger to the heating of the central star and thus give rise to much
larger loops in the \iras\ \cc\ diagram. \\ Our knowledge of dust
opacities, and certainly of solid state features in the mid- and
far-infrared, will improve when the results of the \iso\ mission have been
digested (Waters et al.\q\ 1996).  For example, the well-known 21~\mic\ and
30~\mic\ features that have been observed in the infrared spectrum of
carbon-rich post-AGB stars and planetary nebulae (e.g.\ Omont et al.\q\
1995) have not been taken into account in this study.  In addition, the
wavelength coverage up to 200~\mic\ will be of great help to determine the
wavelength dependence of the dust opacities towards long wavelengths.
\item
A third decisive factor that governs the evolution of the \iras\ \cols\ is
the definition of the end of the AGB. The sooner an object enters the
transition phase, i.e.\ the sooner the heavy AGB mass loss ceases, the longer
the evolution to higher effective temperatures will last.  This results in
cool circumstellar dust shells, and consequently these models are on a
location in the \iras\ \cc\ diagram where not many post-AGB stars were
expected previously. In this respect it is noteworthy to refer to
\fig~\ref{run5} where oxygen-rich post-AGB stars are predicted to be
present in the upper part of region VIII.  A re-investigation of the
sources in that region would be useful in testing whether this scenario is
realistic.
\item
Changing the mass loss prescription in the coolest part of the AGB
evolution that we considered has little influence on the \iras\ \cols\ of
the models.
\end{enumerate}

In general we find that the variation of the parameters mentioned above,
which are still not very well determined, result in a variety of different
paths in the \iras\ \cc\ diagram. This is certainly part of the explanation
why planetary nebulae do not occupy a well-structured region in the \iras\
\cc\ diagram (cf. Volk\q\ 1992).  As a by-product of this investigation we
find that the same location in the \iras\ \cc\ diagram can be occupied by
objects with an entirely different evolutionary past. Apparently, the
location in the \iras\ \cc\ diagram can not {\em a priori} give a unique
determination of the evolutionary status of an object.

As an alternative to the \iras\ \cc\ diagram, the \coli\ vs. \colii\ \col\
diagram is presented. The tracks in this diagram seem less affected by
particulars of the grain emission. Hence this diagram might prove to be a
valuable additional tool for studying post-AGB evolution.

The feedback of observational work, after a sufficient number of parameter
studies, will be of help to assess the selection effects in our post-AGB
sample selection criteria, and will give constraints on the assumptions
that now have to be made on the central star and nebular evolution.

\section*{Acknowledgments}
We are very grateful to Thomas Bl\"ocker who has kindly provided us with
the tables from which the evolutionary timescales could be reconstructed.
The photo-ionization code \cld\ was used. The code is written by Gary
Ferland and obtained from the University of Kentucky, USA.
We would like to thank the referee Ryszard Szczerba and Thomas Bl\"ocker
for critically reading the manuscript.
PvH and RDO were supported by NFRA grants 782--372--033 and 782--372--031.

\end{document}